\documentclass[a4paper,11pt]{article}
\pdfoutput=1 

\usepackage{jheppub} 

\usepackage{latexsym,amssymb,amstext,amsmath}
\usepackage{hyperref}
\usepackage{graphicx} 
\usepackage{subcaption}
\usepackage{amsthm}
\usepackage{esint}
\usepackage{float}
\usepackage{color}

\def\be {\begin{equation}}
\def\ee {\end{equation}}
\def\bea {\begin{eqnarray}}
\def\eea {\end{eqnarray}}

\def\beq{\begin{equation}}
\def\eeq{\end{equation}}
\def\beqa{\begin{eqnarray}}
\def\eeqa{\end{eqnarray}}

\newcommand{\bm}[4]{\begin{pmatrix}#1 & #2\\ #3 & #4 \end{pmatrix}}

\begin{document}
%
\begin{titlepage}

\begin{center}
 {\LARGE\bfseries 
Arithmetic of decay walls \vskip 3mm 
through continued fractions:
 \vskip 3mm
 a new exact dyon counting solution \vskip 3mm 
 in $\mathcal{N} =4$ CHL models}
 \\[10mm]

\textbf{Gabriel Lopes Cardoso, Suresh Nampuri  and Mart\'i Rossell\'o}

\vskip 6mm
{\em  Center for Mathematical Analysis, Geometry and Dynamical Systems,\\
  Department of Mathematics, 
  Instituto Superior T\'ecnico,\\ Universidade de Lisboa,
  Av. Rovisco Pais, 1049-001 Lisboa, Portugal}\\
\vskip 3mm

{\tt gabriel.lopes.cardoso@tecnico.ulisboa.pt}
,\;\,{\tt  nampuri@gmail.com},
\\
{\tt martirossello@tecnico.ulisboa.pt }
\end{center}

\vskip .2in
\begin{center} {\bf ABSTRACT } \end{center}
\begin{quotation}\noindent

\vskip 3mm
\noindent
We use continued fractions to perform a systematic and explicit characterization of the decays of two-centred dyonic black holes in 
$4D$ $\mathcal{N} =4$ heterotic $\mathbb{Z}_N$ CHL models. Thereby we give a new exact solution for the 
problem of counting decadent dyons in these models.

\end{quotation}
\vfill
\today
\end{titlepage}

\tableofcontents

\section{Introduction}

Any candidate theory of quantum gravity
must be able to characterize space-time geometry 
in terms of integral degeneracies of states in the spectrum of the quantum theory.
Further, transitions from one geometry to another must be characterized by changes in these integral degeneracies.
String theory has been shown
to fulfil these expectations, albeit for a restricted class of gravitational systems. Specifically, it has been able to provide  explicit counting formulas for a class of BPS configurations which appear as  supersymmetric black holes in various 
five- and 
four-dimensional theories of gravity \cite{Strominger:1996sh,Dijkgraaf:1996it,Maldacena:1999bp}. In these examples, generating functions for the exact integral statistical degeneracies of BPS microstates underlying a given BPS black hole macrostate have been written down. 
In this note, we will be focusing on $1/4$ BPS  states in four-dimensional $\mathcal{N}=4$ string theory models obtained by compactifying
 heterotic string theory on ${T}^6$, or equivalently Type II on $K3 \times T^2$, as well as its $\mathbb{Z}_N$ orbifold theories called CHL models,
 with $N=2,3,5,7$ \cite{Chaudhuri:1995fk, Chaudhuri:1995bf, Schwarz:1995bj, Vafa:1995gm, Chaudhuri:1995dj, Dijkgraaf:1996it, Jatkar:2005bh, David:2006ji, David:2006yn, Sen:2007vb, Dabholkar:2007vk,  Cheng:2007ch, Banerjee:2008pv, Dabholkar:2008tm, Banerjee:2008pu, 
Dabholkar:2008zy, Cheng:2008fc, Sen_2011,  Dabholkar:2012nd, Ferrari:2017msn, Paquette:2017gmb, Bossard:2018rlt, Chattopadhyaya:2018xvg, Chowdhury:2019mnb, Fischbach:2020bji}.
For simplicity,  we will refer to all these theories, including the parent theory, as CHL theories, regarding  the parent heterotic theory as corresponding to a trivial $\mathbb{Z}_1$ orbifold. 
 
 A generic  BPS configuration in these theories carries electric and magnetic charges with respect to multiple Abelian $U(1)$ gauge fields and hence is referred to as a dyonic BPS state. $1/4$ BPS generating functions in these theories have been written down in terms of Siegel modular forms defined on a $\mathrm{Sp}(2,\mathbb{Z})$ upper half plane, called Siegel upper half plane. 
 These generating functions count three kinds of $1/4$ BPS dyonic states, namely:
those that gravitate to form single centre $1/4$ BPS dyonic black holes with a finite horizon area in two-derivative gravity,  
two-centred bound states of $1/2$ BPS constituents, and finally single centre states that have zero horizon area at the two-derivative level. 
The single centre states exist in all regions of moduli space and are hence labelled immortal.
For each two-centred bound dyon, there exist co-dimension one loci in the moduli space of the theory across which the said state marginally decays into its 
$1/2$ BPS 
constituents. Consequently, these states are called decadent dyons, while the corresponding decadent loci are referred to as  their 
decay walls. We will be studying dyonic decays in a two-dimensional space, called the axion-dilaton moduli space, where the decay walls 
correspond to
 lines of marginal stability (LMS).
 A  decadent dyon, therefore, exists only on one side of its line of marginal stability.  The generating function has an infinite family of second order poles, 
 a subset of which are parametrized by matrices in\footnote{$\Gamma_0(N) = \left\{ \bm{a}{b}{c}{d} \in  {\rm PSL} (2, \mathbb{Z}) \; \vert \;
   c = 0 \mod  N \right\} \;. $ } 
 $\Gamma_0(N) \subset {\rm PSL} (2, \mathbb{Z})$
  and correspond to the decadent lines of marginal stability. The appearance or disappearance of two-centred bound states  when crossing a line of marginal stability changes the degeneracy by an amount equal to the residue of the generating function at the corresponding pole. This change is referred to as a  wall-crossing jump $(\Delta d)_{\text{LMS}}$. The residue contribution at each pole can be computed by mapping it
 to a single `canonical' pole which corresponds to the simplest possible dyonic decay, 
 namely the dyon splitting into a $1/2$ BPS electric and a $1/2$ BPS magnetic monopole. The wall-crossing jump then is simply the product of the known degeneracy of each constituent and their electromagnetic angular momentum, called the Dirac-Schwinger-Zwanziger
angular momentum. 
 The existence of lines of marginal stability guarantees that there exists a region of moduli space, where a given decadent 
 dyon has decayed and no longer exists as a two-centred system. It 
    completely disappears from the $1/4$ BPS spectrum and does not contribute to the dyonic degeneracy in this region.
 Hence, decadent dyonic degeneracy at a given point in moduli space can be calculated as follows. Starting from this point, as one moves through moduli space to the terminal region where the dyonic degeneracy is known or independently determined, 
 a given decadent dyon can decay across one  of a series of 
lines of marginal stability corresponding to possible decay modes. 
The total change in dyonic degeneracy evaluated as the sum of the wall-crossing jumps at these decay lines is the difference of the  degeneracies at its initial and terminal points. Hence the decadent dyonic degeneracy in the initial region is 
\begin{equation}\label{first}
    d_{\text{Initial}}\,=\,d_{\text{Terminal}}\,+\,\sum_{\text{LMS}} (\Delta d)_{\text{LMS}} \;.
\end{equation} 
To exactly compute this change,
we need a systematic method of tracking and characterizing the lines of marginal stability encountered by a given dyon as it moves to a terminal region, where its degeneracy contribution is either zero or a finite quantity that is independently determined.  For a given decadent dyon it has been shown \cite{Sen:2007vb,Sen_2011}
 that these can be encoded as a set ${W}$ of $\Gamma_0(N)$ matrices.\footnote{For an analysis of the walls of marginal stability in $\mathcal{N}=2$ string theories, see \cite{David:2009ru}.}
  The matrices in this set were recently  studied  by \cite{Chowdhury:2019mnb}, who showed that the set was finite and translated the counting problem into constraints on these matrices. In the absence of an explicit characterization of the set ${W}$, their analysis requires careful identification of certain  decay modes corresponding to black hole bound state metamorphosis (BSM) \cite{Sen_2011,Chowdhury:2012jq,Chowdhury:2019mnb} in order to avoid overcounting.  

In this note, we will provide a systematic and explicit characterization of this set  ${W}$ by setting up a
new approach based on continued fractions for decadent dyons.
This arithmetic encodes a rule for generating elements of ${W}$, using the continued fraction representation of a ratio of two charge bilinears associated with the decadent dyon.
This new approach automatically avoids all overcounting complications, such as BSM, associated with dyonic decays.
The basic principles of this new decadent dyon counting framework can be summed up as follows: 
\begin{enumerate}

\item Decadent dyons decay across lines of marginal stability in moduli space, either to disappear from the $1/4$ BPS spectrum or to leave behind an immortal remnant with a vanishing classical horizon and with a finite contribution to the BPS degeneracy that is independently determined.
Hence,  for every decadent dyon, there exists a terminal region, i.e., a region of moduli space where its degeneracy is either vanishing or finite and known.

\item 
The lines of marginal stability are in a one-to-one correspondence 
with a subset of the second order poles of the generating function, parametrized by the group $\Gamma_0(N)$.\footnote{In the $N=1$ case, the lines of marginal stability are in 
one-to-one correspondence with $\Gamma_+(1)$, see \eqref{symmetry}.}

\item The lines of marginal stability divide the axion-dilaton moduli space into
connected regions/chambers 
\cite{Sen_2011}. Following \cite{Chowdhury:2019mnb}, we will compute dyonic degeneracies in one of these regions, called the $\mathcal{R}$-chamber.

\item The lines of marginal stability that a given dyon encounters in its trajectory through the axion-dilaton moduli space from the $\mathcal{R}$-chamber to its terminal region,  where its degeneracy is known or independently determined, are encoded in a set ${W}$ of matrices in  $\Gamma_0(N)$.

\item Our decadent dyon counting principle can be stated as follows:

{\it The sequence of lines of marginal stability  which encode   decadent dyonic degeneracy is systematically generated by the continued fraction representation of two integers associated with the decadent dyon}. \\

\end{enumerate}

The salient features of the results of this paper are briefly outlined as follows.

We solve the decadent dyon counting problem specified in terms of three charge bilinears $(m,\,n,\,\ell)$ with
$\Delta = 4 m n - \ell^2 \leq 0$.
Using the observation that the lines of marginal stability are Farey arcs which correspond to $\Gamma_0(N)$ matrices, we show that the finite sequence of $\Gamma_0(N)$ matrices in $W$ has the distinctive property that its last element, which we denote by  $\gamma_*$, completely determines $W$.

Using known constraints on the charge bilinears to fix $\gamma_*$, each element of $W$ can be read off from its decomposition in terms of the
$\mathrm{PSL} (2, \mathbb{Z})$ 
 basis $\{\bm{1}{1}{0}{1},\,\bm{1}{0}{1}{1}\}$.
  This fixes $W$ explicitly. Hence, the formula \eqref{first} can be used to compute the dyonic degeneracy $d_{\text{Initial}} (m,\,n,\,\ell)$. For $\Delta =0$, we find a  remarkable property for $d_{\text{Terminal}}$: it is the dyonic degeneracy of an immortal dyonic configuration with charge bilinears $(0, \frac1N \gcd(m, n N, \ell), 0)$ 
and captured by a known mock modular form \cite{Dabholkar:2012nd,Bossard:2018rlt} for each CHL orbifold.

This note is structured as follows. In section \ref{sec:dc} we will define the dyon counting problem by first reviewing the setup and notation employed in characterizing dyons and dyonic degeneracies in $\mathcal{N}=4$ CHL  theories and, in particular, explicate the representation of each line of marginal stability by a $\Gamma_0(N)$ matrix. The review material in this section follows \cite{Sen_2011}. In section \ref{sec:ddcp}, we will solve the dyon counting problem by introducing relevants aspects of the theory of continued fractions, and we will explicitly show how the requisite lines of marginal stability for computing a 
decadent dyonic degeneracy are generated by a continued fraction representation. Our solution for characterizing the lines of marginal stability can be displayed as an elegant diagrammatic representation of decay walls based on continued fraction convergents and Farey arcs. In 
section \ref{sec:disc} we systematize various features of the reasoning behind and implications of our results for $\mathbb{Z}_N$ CHL models.
In the appendix we comment on the relation of our continued fraction approach to the theory of integral binary quadratic forms and comment on BSM.

\section{Dyon counting problem \label{sec:dc}}

\subsection{Notation and setup \label{sec:notset}}

We consider $\mathbb{Z}_N$ CHL orbifold models with $N=1,2,3,5,7$.  These models have $r= 48/(N+1)+4$ Abelian gauge fields with respect to
which a generic BPS state in these theories carries electric and magnetic charges \cite{Jatkar:2005bh}. These form vectors $\Vec{Q}$ and $\vec{P}$  respectively. $\vec{P}$ lies in an $r$-dimensional even lattice that for any $N>1$ model is not self-dual, while $\Vec{Q}$  lies in the dual lattice to $\vec{P}$ such that $N\Vec{Q}$  itself belongs to an $r$-dimensional  integral even lattice.  Dyonic degeneracies are functions of three rational numbers associated with the dyons, namely: the norms of the electric and magnetic vectors, $Q^2=2 n$ and $P^2 =2m$ respectively, and the inner product of the electric and magnetic vectors, $Q \cdot {P} =\ell$, which form the T-duality invariants of the theory. 
Further, the $1/4$-BPS states in these theories are characterized in terms of two 
U-duality invariants \cite{Dabholkar:2007vk,  Dabholkar:2008tm,Dabholkar:2008zy,Sen:2009vz},
\begin{equation}
    \Delta =Q^2\,P^2\,-\,(Q\cdot P)^2 = 4 mn - \ell^2 \;,
\end{equation}
\begin{equation}
    I = \gcd(Q_i\,P_j\,-\,Q_j\,P_i) \;\;\;,\;\;\; 1\,\leq\,i,\,j\,\leq\,r \;,
\end{equation}
where $Q_i$ and $P_j$ are the $i^{\text{th}}$ and $j^{\text{th}}$ components of $\vec{Q}$ and $\vec{P}$.  
For $N=1$, 
the dyonic degeneracies for arbitrary 
torsion $I$ are explicitly given in terms of the degeneracies $d(m, n, \ell)$ for $I=1$ as \cite{Banerjee:2008pu,Dabholkar:2008zy}
\begin{equation}
d_I(m, n, \ell)\,=\,\sum_{t|I} t \; d(m,\,\frac{n}{t^2},\,\frac{\ell}{t}) \;.
\end{equation}
In the following, we will focus on the $I=1$ case for $N=1,2,3,5,7$. Later, we will generalize our results to $I >1$, but $N=1$ alone.

Single centre $1/4$ BPS black holes with finite horizon area have $\Delta >0$. In this paper, we will consider
two-centred $1/4$ BPS configurations with $\Delta \leq 0$. For the case $N=1$, it was shown in 
\cite{Ferrari:2017msn} that 
the degeneracies of
single centre  $1/4$ BPS black holes are encoded in states with $\Delta < 0$ through a generalized Rademacher expansion.

\begin{figure}[t!]
	\centering
		\includegraphics[scale=0.4]{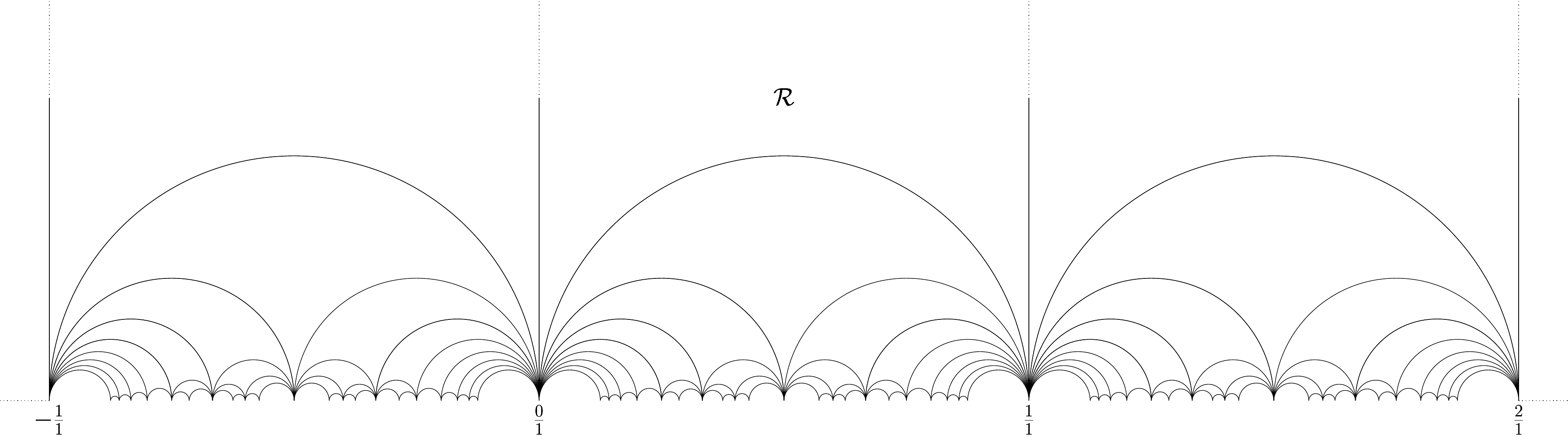}
	\caption{Lines of marginal stability in the axion-dilaton complex upper half plane for the case $N=1$.}
	\label{fareyseq}
\end{figure}

We will focus on 1/4 BPS states with primitive charges, and belonging to the twisted sector of the $\mathbb{Z}_N$ CHL orbifold
when $N > 1$.
The generating function for dyonic degeneracies of these 1/4 BPS states\footnote{See 
\cite{Banerjee:2008pv,Bossard:2018rlt,Fischbach:2020bji} for BPS degeneracies in the untwisted sector.} 
is a
modular form\footnote{ $\Phi_k$ is a modular form of weight $k = 24/(N+1) -2$ under a subgroup ${\tilde G} \subset \mathrm{Sp} (2, \mathbb{Z})$ \cite{Jatkar:2005bh}. It
transforms as 
 \begin{equation} \Phi_{k}(\Omega)\,\rightarrow\, \det (C\,\Omega + D)^k\,\Phi_k (\Omega)
,\nonumber\end{equation} where $\Omega =\bm{\rho}{v}{v}{\sigma}$ denotes the period matrix of the genus-2 Riemann surface which
transforms as $\Omega\,\rightarrow\, (A\,\Omega\,+\,B )(C\,\Omega\,+\,D)^{-1} $ under transformations
 $\bm {A}{B}{C}{D} \in {\tilde G}$.}
 of a subgroup of the genus-2 modular group $\mathrm{Sp} (2, \mathbb{Z})$ \cite{Jatkar:2005bh}, (we use the conventions of \cite{Sen_2011})
\begin{equation}\label{1}
    \frac{1}{\Phi_k (\rho, \sigma, v)} = \sum_{\begin{matrix}m,nN \geq  -1\\ m, nN, \ell \in \mathbb{Z}\end{matrix}} 
    (-1)^{\ell +1} \, 
    d(m, n,\ell)\, e^{2 \pi i \,(m \rho + n \sigma + \ell   v )}.
\end{equation}
$\Phi_k$ has an infinite family of second order zeroes in the $(\rho, \sigma, v)$ Siegel upper half plane. A subset of these satisfy (see eq. (5.1) in \cite{Sen:2007vb}; we follow the notation of \cite{Sen_2011} )
\begin{equation}
p q \sigma_2\,+r s \rho_2 +(ps +q r) v_2\,=\,0 \;\;\;,\;\;\;   \bm{p}{q}{r}{s} \in \Gamma_0(N) \;,
\label{loci0}
\end{equation}
where  $(\rho_2,\,\sigma_2, v_2)$ are the imaginary parts of $(\rho,\,\sigma , v)$,  and where
\bea
\Gamma_0(N) = \left\{ \bm{a}{b}{c}{d} \in  {\rm PSL} (2, \mathbb{Z}) \; \vert \;
   c = 0 \mod  N \right\} \;. 
   \label{gam0}
   \eea
These loci form co-dimension 1 hypersurfaces or walls which delineate different domains in the  $(v_2/\sigma_2,\, \rho_2/\sigma_2) $ plane. These correspond to the lines of marginal stability in the axion-dilaton complex upper half plane \cite{Sen:2007vb,Sen_2011}, as shown in Figure \ref{fareyseq}.

At each of these lines, two-centred dyons can fragment  into individual $1/2$ BPS constituents and disappear from the spectrum. 
The  wall-crossing jump is precisely the residue of the generating function evaluated at the corresponding pole. 
Each of these lines falls into one of two categories \cite{Sen_2011}: they are either circles that intersect the horizontal axis at two rational points,  $p/r$ and $q/s$, with $p s - q r = 1$, or they are 
vertical lines at integral points on the horizontal axis. The latter can be viewed as special circles that connect $p/r$ and $q/s$, with vanishing
$r$ or $s$. We will refer to all of these circles as Farey arcs.
 Hence, associated with each line of marginal stability is a $\Gamma_0(N)$ matrix $\bm{p}{q}{r}{s}$, 
formed from its rational endpoints  $p/r$ and $q/s$.
The $i^{\text{th}}$ vertical line lies $i$ units to the right of the vertical axis and is labelled by the $i^{\text{th}}$ power of the $T$-generator of $\mathrm{PSL}(2,\mathbb{Z})$ given by $\bm{1}{1}{0}{1}$. Hence, vertical lines are referred to as T-walls and divide the axion-dilaton  complex upper half plane into semi-infinite strips, each of which contains an infinite number of Farey arcs. In particular, the T-wall passing through the origin of the 
axion-dilaton  complex upper half plane corresponds to the unit matrix.  
The lines of marginal stability represented by circles with finite rational points  $p/r$ and $q/s$ are called S-walls.

The decay modes at each line of marginal stability are completely determined by the corresponding $\Gamma_0(N)$ matrix, as is the corresponding zero \eqref{loci0}  of $\Phi_k$, as \cite{Sen_2011}
\begin{equation}\label{gensplit}
\gamma\,=\,\bm{p}{q}{r}{s}\,\in\, \Gamma_0(N):\, \begin{pmatrix}Q\\P\end{pmatrix}\,\rightarrow\,\begin{pmatrix} p \,(s\,Q\,-\,q\,P)\\r\,(s\,Q\,-\,q\,P)\end{pmatrix}\,+\,\begin{pmatrix}q\,(- r \,Q\,+\,p\,P)\\ s \,(-  r \,Q\,+\,p\,P)\end{pmatrix} \;.
\end{equation}
The decay mode corresponding to the unit matrix is the simplest case of a dyonic decay: the dyon undergoes an `elementary' split into a pure electric and a pure magnetic fragment, as can be seen by putting $q=r=0$ and $p=s=1$ in the above formula to get 
\begin{equation}\label{minsplit}
\gamma\,=\,\bm{1}{0}{0}{1}\,:\, \begin{pmatrix}Q\\P\end{pmatrix}\,\rightarrow\,\begin{pmatrix}Q\\0 \end{pmatrix}\,+\,\begin{pmatrix}0\\P\end{pmatrix},\,\,\, 
v_2 \,=\,0
\;.
\end{equation}
We will refer to this T-wall as an elementary T-wall.
The corresponding wall-crossing jump produces a change in the dyonic degeneracy formula. 
This change can be computed
by observing that for $v\,\rightarrow\,0$ \cite{Jatkar:2005bh}, 
\begin{equation} \Phi_k(\rho, \sigma,\,v)\,\xrightarrow{v\,\rightarrow\,0} \,v^2\,   f^{(k)} (\rho)\,f^{(k)} (\frac{\sigma}{N}) ,
\label{Pffk}
\end{equation}
where $f^{(k)} $ is a specific weight $(k+2)$ modular form of $\Gamma_0(N)$,  namely \cite{Jatkar:2005bh} $f^{(k)} (\rho) = \eta (\rho)^{k+2} \,  \eta (N \rho)^{k+2} $, where $\eta(\rho)$ denotes the Dedekind eta function. We write the Fourier expansion of its inverse as 
\be
\frac{1}{f^{(k)} (\rho)}  
= \sum_{m=-1}^{\infty} d_1 (m)  \,e^{2 \pi i m \rho}  \;\;\;,\;\;\;
\frac{1}{f^{(k)} (\sigma/N)}  
= \sum_{n=-1/N}^{\infty} d_2 (n)  \,e^{2 \pi i n \sigma}  \;.
\label{d1d2}
\ee
Hence, the wall-crossing jump across this T-wall can be deduced  from the $v=0$ residue of the generating function to be \cite{Sen_2011}
\begin{equation}
    \Delta\,d(m, n, \ell)\,=\, (-1)^{\ell\,+\,1}\,|\ell|\, d_1(m)\, d_2(n) \;.
\end{equation}

We now make an important point about encoding the orientation of the decay walls in terms of a $\Gamma_0(N)$ matrix $\bm{p}{q}{r}{s}$ \cite{Sen_2011}.
We take the orientation to run from the second column to the first column, which represents the wall running from $q/s$ to $p/r$. Hence,
the elementary T-wall runs from $0$ to $ i \infty$.
It can be shown \cite{Sen_2011} that for $\ell > 0$, the two-centred state $(m, n, \ell)$ exists to the left of the elementary T-wall and disappears from the $1/4$ BPS spectrum  as one crosses the wall from left to right. Conversely, the $\ell < 0$ two-centred state exists to the right of this wall and decays across it as we move from right to left.  
 The labelling of lines of marginal stability by $\Gamma_0(N)$ matrices allows us to map a generic dyon decay as in \eqref{gensplit} to the elementary 
 T-wall decay, 
 \begin{equation}\label{frag}
    \gamma^{-1}\ \begin{pmatrix}Q\\P\end{pmatrix} = \begin{pmatrix}Q_{\gamma} \\P_{\gamma} \end{pmatrix}
        \,\rightarrow\,\begin{pmatrix} s Q - qP \\0\end{pmatrix}\,+\,\begin{pmatrix}0\\ - r Q\,+ pP\end{pmatrix}\,=\, \begin{pmatrix}Q_{\gamma}\\0\end{pmatrix}\,+\,\begin{pmatrix}0\\P_{\gamma}\end{pmatrix} \;.
\end{equation}
The charge bilinears $(m, n, \ell)$ are mapped to $(m_{\gamma}, n_{\gamma}, \ell_{\gamma})$,
\begin{eqnarray}
\frac{Q^2_\gamma}{2} &=& n_\gamma = s^2\, n + q^2\,m - q s \,\ell \; ,\nonumber\\
\frac{P^2_\gamma}{2} &=& m_\gamma = r^2\,  n +\, p^2\,m - p r \,\ell \;, \nonumber\\
Q_\gamma \cdot  P_\gamma &=& \ell_\gamma =- 2 r s \, n\,- 2 p q \,m\,+( p s + q r )\,\ell \;. 
\label{frag3}
\end{eqnarray}
Under this map, a two-centred $1/4$ BPS state $(m,n,\ell)$ which exists to the left 
of a Farey arc, and decays across it,
is mapped to a two-centred $1/4$ BPS state $(m_{\gamma},n_{\gamma},\ell_{\gamma}>0)$ which exists to the 
  left of the elementary T-wall,
 while a two-centred $1/4$ BPS state $(m,n,\ell)$ which exists to the right
of a Farey arc, and decays across it,
is mapped to a two-centred $1/4$ BPS state $(m_{\gamma},n_{\gamma},\ell_{\gamma}<0)$ 
which exists to the right of the elementary T-wall.

{\it  Thus, the wall-crossing jump contribution of a dyon $\begin{pmatrix}Q\\P\end{pmatrix}$ across a generic line of marginal stability, labelled by a $\Gamma_0(N)$ matrix $\gamma$, to the dyonic degeneracy formula is equal to the jump contribution of the dyon $\begin{pmatrix}Q_\gamma\\P_\gamma\end{pmatrix}$ across the elementary T-wall} 
\cite{Sen_2011},
\begin{equation}\label{deca}
\Delta_\gamma \,  d(m,\,n,\,\ell) = (-1)^{\ell_\gamma +1}\,| \ell_{\gamma} | \, d_1(m_\gamma)\,d_2(n_\gamma) \;.
\end{equation} 
Hence, counting contributions from the decay of a given dyon across various lines of marginal stability  is equivalent to counting decays for corresponding $\gamma$-equivalent dyons across the elementary T-wall.  

Further, given a Farey arc with $r s> 0$  and $p s - q r = 1$, the relation $\frac{p}{r}-\frac{q}{s} = \frac{
 1}{r s}$  implies that the second column rational number is less than the first column one. If we reverse the orientation of the $\gamma$- matrix by switching the two columns by making use of the $S$-generator of $\mathrm{PSL}(2, \mathbb{Z})$ given by $\bm{0}{-1}{1}{0}$,
  \begin{equation}
    \gamma\,\rightarrow\, \gamma \bm {0}{-1}{1}{0} \;,
 \end{equation}
then the corresponding $\ell_\gamma$ flips sign and $r\,s\,<\,0$. The effect of the column switching is to interchange the two fragments in the $\gamma$-frame i.e. the elementary split now looks like 
\begin{equation}
    \begin{pmatrix}Q_\gamma\\P_\gamma\end{pmatrix}\,\rightarrow\,\begin{pmatrix}0\\P_\gamma\end{pmatrix}\,+\,\begin{pmatrix}Q_\gamma\\0\end{pmatrix}
    \;.
\end{equation}
Following \cite{Chowdhury:2019mnb}
we now define  
\begin{eqnarray}
    \Gamma_+(N) &=& \left\{\gamma\,=\,\bm{p}{q}{r}{s} \,\in \, \Gamma_0(N)|\, rs > 0\right\}, \nonumber\\
  \Gamma_-(N) & =&  \left\{\gamma\,=\,\bm{p}{q}{r}{s} \,\in  \Gamma_0(N)|\, rs < 0 \right\}.
\end{eqnarray}
We note the relation
 \begin{equation}\label{symmetry}
    \Gamma_-(1) =  \Gamma_+(1)\,\bm{0}{-1}{1}{0} = \Gamma_+(1)\,S \;.
\end{equation}
{From} the above discussion we conclude that lines of marginal stability labelled by matrices in $\Gamma_{+}(N)$ with $\ell_\gamma >0 $, 
and in $\Gamma_{-}(N)$ with $\ell_\gamma <0 $,
correspond to dyonic decays where the decadent dyon exists above the lines of marginal stability.

\subsection{Defining the dyon counting problem }

We now define the decadent dyon counting problem for CHL dyons with $I=1$.

 Consider the ${R}$-strip sandwiched between the elementary T-wall at $0$ and 
 the T-wall at $1$.  The $\mathcal{R}$-chamber is the region in the ${R}$-strip adjoining
the elementary T-wall and which does not contain lines of marginal stability.
  We consider a decadent two-centred state with $\Delta \leq 0$
 that exists in a region of the axion-dilaton moduli space corresponding to the  $\mathcal{R}$-chamber. 
 It will decay along some loci in the upper half plane\footnote{The largest Farey arc in this strip is the semi-circle intersecting the horizontal axis at $0$ and $1$, and it is a decay wall only for the $N=1$ case.}.
 Starting in the  $\mathcal{R}$-chamber, as we move down in the  ${R}$-strip, the dyon encounters a progression of  decay walls 
 represented by increasingly smaller Farey arcs.
Each such arc can be mapped to a $\gamma$-frame corresponding to an elementary split of a $\begin{pmatrix}Q_\gamma\\P_\gamma\end{pmatrix}$. 
If
$\gamma$ lies in $\Gamma_+(N)$  (in $\Gamma_-(N)$) and 
 the corresponding $\ell_\gamma$ satisfies $\ell_\gamma > 0$ ($\ell_\gamma < 0$), then 
 by the arguments given above, the decadent dyon will exist above the Farey arc and decay across it. To put it another way, starting from a chosen point in the  ${R}$-strip lying below a particular Farey arc, any trajectory directed vertically upwards in the  $\mathcal{R}$-chamber will encounter a unique finite sequence of increasingly larger Farey arcs till it enters the  $\mathcal{R}$-chamber.

 Therefore, given a 
 point in the  ${R}$-strip lying below the  $\mathcal{R}$-chamber, one  can construct a sequence of $\Gamma_0(N)$ matrices
 corresponding to the sequence of Farey arcs that will be intersected by a vertical line starting from this point and ending in the  $\mathcal{R}$-chamber.
 In this note, given a triplet $(m, n,\ell)$,  we will explicitly construct a sequence $W (m, n, \ell)$ 
 to compute the dyonic degeneracy $d(m,n,\ell) $ in the  $\mathcal{R}$-chamber 
   via the master formula \eqref{truemaster} given below, 
while avoiding overcounting complications such as BSM.

 The elements in the sequence $W (m, n, \ell)$ will be indexed by natural numbers in the order in which they are encountered as one moves down from the $\mathcal{R}$-chamber.\footnote {That is, the largest Farey arc $\gamma$ matrix will be $\gamma_1$, the next smaller one will be $\gamma_2$ and so on.}  
 If $d_*$ represents the known or independently computable
 dyonic degeneracy at a point $*$ in the strip, separated from the $\mathcal{R}$-chamber by $k$ walls (Farey arcs) corresponding to the 
 $k$ elements of $W (m, n, \ell)$, with the wall-crossing jump across the $i^{\text{th}}$ wall denoted by $\Delta_i$, then the master equation for 
 computing decadent dyonic degeneracies in the $\mathcal{R}$-chamber is given by\footnote{From \eqref{frag3}, 
 $\ell_\gamma = \ell \mod 2$, allowing us to replace $(-1)^{\ell_{\gamma}+1}$
  in \eqref{deca} by 
 $(-1)^{\ell +1}$.} 
\begin{equation}\label{truemaster}
d(m, n, \ell) = d_*
+  \sum_{i=1}^k\Delta_i = d_* + (-1)^{\ell+1} \sum_{\substack{i =1\\ \gamma_i\,\in\, W (m, n, \ell) }}^k\,
| \ell_{\gamma_i}|  \, 
d_1(m_{\gamma_i})\,d_2(n_{\gamma_i})\;,
\end{equation}
where $d_1$ and $d_2$ where introduced in \eqref{d1d2}.

{From} \eqref{1}, it is clear that if the $k^{\text{th}}$ $\gamma$-frame charges satisfy $m_{\gamma_k}<-1$ or $n_{\gamma_k} <-1/N$, then the dyonic degeneracy $d(m_{\gamma_k},n_{\gamma_k},\ell_{\gamma_k})$ 
is zero.
 In this case, the total change in the dyonic degeneracy due to wall-crossing jumps across the sequence $W (m, n, \ell)$, starting from the largest Farey arc to the last wall below where it ceases to exist, is equal in magnitude to the degeneracy of the dyon in the $\mathcal{R}$-chamber. 
  Hence, given 
  $(m, n, \ell)$ for which one can identify a finite decay sequence $W (m, n, \ell)$ of $\Gamma_0(N)$ matrices such that the endpoint 
  $\gamma_*$ of this sequence corresponds to a charge configuration with vanishing contribution 
  $d_*$ to the dyon counting formula, 
  then the decadent dyonic degeneracy in the 
  $\mathcal{R}$-chamber can be written as \cite{Sen_2011,Chowdhury:2019mnb}
 \begin{equation}\label{nmaster}
 d(m, n, \ell)\,=  (-1)^{\ell + 1} \, \sum_{\gamma\,\in W (m, n, \ell)} \, | \ell_\gamma | \, d_1(m_{\gamma})\,d_2(n_{\gamma}) \;.
 \end{equation}

 We are now ready to give an operational definition of the dyon counting problem as follows: 
given charge bilinear invariants $(m, n, \ell)$ with $\Delta \leq 0$, 
\begin{enumerate}
    \item define a finite dyonic decay sequence $W(m, n, \ell)$ of $\Gamma_0(N)$ matrices 
    corresponding to decay walls in the ${R}$-strip, such that the endpoint of this sequence corresponds to  bilinears $(m_\gamma, n_\gamma,\ell_\gamma)$ with known or independently computable dyonic degeneracy $d_* = d(m_\gamma, n_\gamma,\ell_\gamma)$.
    \item Then use the master formula \eqref{truemaster} to compute $d(m, n, \ell)$, the dyonic degeneracy in the $\mathcal{R}$-chamber. 
    \end{enumerate}
    
 \section{Solving the decadent dyon counting problem \label{sec:ddcp}}
 
 \subsection{Dyonic decay sequence $W(m,n,\ell)$}
 
 In order to characterize  $W(m,n,\ell)$, we will first restrict our analysis to
 decadent dyons with torsion $I=1$ in
  the heterotic string theory on $T^6$, the ${N}\,=\,1$ CHL model. 
  We will consider decadent dyons with $I >1$ in subsection \ref{sec:decd}.
  Our formulation of decadent dyons with $I=1$ will be easily generalized subsequently to the ${N}\,=\,2,\,3,\,5,\,7$ CHL models, as will be discussed
  in subsections \ref{sec:subD0}
   and \ref{sec:CHL}.

 In the heterotic string theory on $T^6$, the lines of marginal stability are labelled by  $\mathrm{PSL}(2,\mathbb{Z})$  matrices,  with 
 $[ f^{(k)} (\rho ) ]^{-1} $ in 
 \eqref{d1d2} given by
  $[\eta^{24} (\rho)]^{-1} =  \sum_{m \in \mathbb{Z}, \,m \geq \,-1} d_1(m) \, e^{2\pi i m \rho}$. 
  Arranging the three T-duality invariant charge bilinears in a  binary quadratic form, 
 \begin{equation}
     A\,=\, \bm{2m}{-\ell}{-\ell}{\; 2n},
 \end{equation}
   the $\mathrm{PSL}(2,\mathbb{Z})$ transformations
   \eqref{frag3}
   on the charges which bring them into the $\gamma$-frame can be represented as a $\gamma$ conjugation operation, 
  \begin{equation}
      \gamma:\, A\,\rightarrow\, \gamma^T \,A\,\gamma = A_\gamma\,=\,\bm {2\,m_\gamma}{-\ell_\gamma}{-\ell_\gamma}{\; 2\,n_\gamma}.
      \label{conjA}
  \end{equation}
Thus, the problem of characterizing $W(m,n,\ell)$ can be re-stated as the problem of identifying a corresponding sequence $\{A_\gamma\}$. 
Recall that the $\gamma$-matrices in $W(m,n,\ell)$ generating this sequence will be  in $\Gamma_{+}(N)$ (respectively $\Gamma_{-}(N)$),
 in which case the corresponding line of marginal stability contributes to the decadent dyon degeneracy if $\ell_\gamma >0$
(respectively  $\ell_\gamma < 0$). 
Further, for every $\gamma_- \in  \Gamma_- (1)$, there exists a $\gamma_+ \in \Gamma_+(1)$ 
such that $\gamma_- = \gamma_+ \,S$. As $S$ is a basis generator of $\mathrm{PSL}(2,\mathbb{Z})$, and hence is a symmetry of
the ${N} =1$ CHL model, 
we can choose the matrices $\gamma$ in  $W(m,n,\ell)$ to lie solely in $\,\Gamma_+(1)$.

 \begin{figure}[t!]
	\centering
		\includegraphics[scale=0.8]{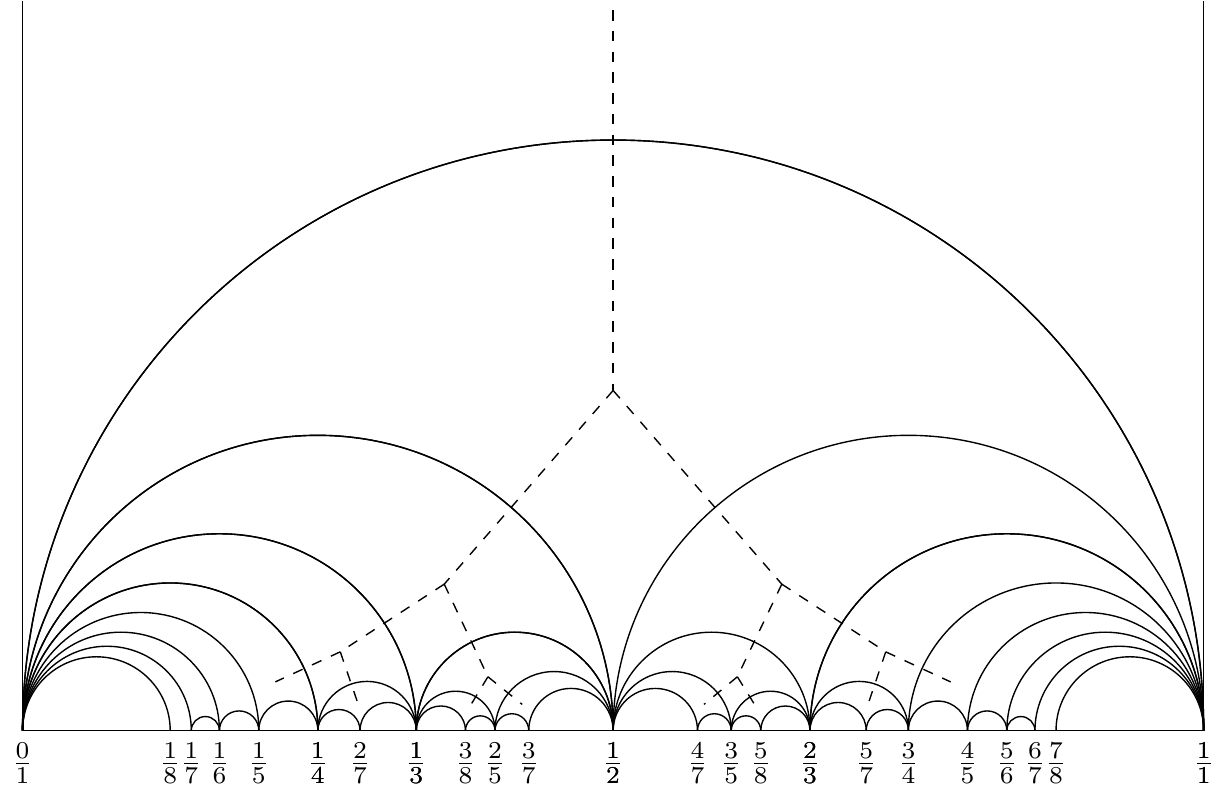}
	\caption{Options for downward decay trajectories.}
	\label{fig:deci}
\end{figure}

In order to solve the dyon counting problem, we must first construct a $W(m,n,\ell)$,  or alternatively a finite Farey arc sequence, such that in the region below the last Farey arc in the sequence, the dyonic degeneracy $d_*$ is known or independently determined. We do this by observing that there exists a simple algorithm to construct the Farey arcs in the $R$-strip starting from the largest and first arc that is encountered as one moves in the ${R}$-strip down from the $\mathcal{R}$-chamber. Figure \ref{fig:deci}
depicts possible options for downward decay trajectories.
The arcs are labelled in the order in which they are encountered. 
Let us denote the region
below the $i^{\text{th}}$ arc by  $\mathcal{R}_i$.  The first arc that is inevitably crossed is the arc between $0$ and $1$ with a radius of $\frac{1}{2}$. This corresponds to $\gamma_1=\bm{1}{0}{1}{1}$. In $\mathcal{R}_1$, there is a binary choice to turn left or right in order to intersect the arc between $0$ and $\frac{1}{2}$ and that between $\frac{1}{2}$ and $1$ respectively, corresponding to matrices $\gamma_2= \bm{1}{0}{2}{1} =\gamma_1^2$ and 
$\gamma'_2  =\gamma_1\,\bm{1}{1}{0}{1}$. This process can be repeated for both $\gamma_2$ and $\gamma'_2$. Hence, at every level, there exists a binary left-right choice, which is equivalent to multiplying the $\gamma$ matrix at that level by either $U =\bm{1}{0}{1}{1}$ 
or $T=\bm{1}{1}{0}{1}$. Hence, 
the latter form a natural choice for a  basis of $\mathrm{PSL}(2,\,\mathbb{Z})$ in terms of which to express decay matrices in $W(m,n,\ell)$. 
Thus, given the basis decomposition of an $\mathrm{PSL}(2,\,\mathbb{Z})$ matrix $\gamma$ in $W(m,n,\ell)$, 
\begin{equation}\label{span}
    \gamma = U^{s_1}\,T^{s_2}\,U^{s_3}\, \cdots\,T^{s_r} \,, \;\;\;\; s_1, \dots ,  s_{r-1}  > 0, \; \; s_r \geq 0  \;,
\end{equation}
the sequence $W(m,n,\ell)$ corresponding to all Farey arcs starting from the largest one till the arc $\gamma_*$ is given as 
\begin{equation}\label{seq}
    W(m,n,\ell) = \left\{U,\,U^2,\, \dots , U^{s_1},\,U^{s_1}\,T,\, \dots , U^{s_1}{T^{s_2}},\,U^{s_1}\,{T^{s_2}}\,U,\, \dots , U^{s_1}\,{T^{s_2}}\,U^{s_3} , 
     \dots,\,\gamma_* \right\}  \;.
\end{equation}

We now proceed to determine $\gamma_*$ for the decadent dyons. We will first determine $\gamma_*$ for $\Delta < 0$.
The case $\Delta =0$ will be analyzed in subsection \ref{sec:del0}.

We consider $\Delta = 4 m n -\,\ell^2 < 0$, and we restrict to  $0 \leq  \ell \leq m $ 
without loss of generality, following \cite{Chowdhury:2019mnb}. Note that this implies $m>0$.
    In this case, as shown in \cite{Sen_2011}, one can always choose a $\gamma_*$  such that 
      \begin{equation}
        m_{\gamma_*} < -1  \;\; \lor  \;\; n_{\gamma_*} < -1 \;\;  \,\,\implies d_* = 0.
    \end{equation}
    For concreteness, we will restrict to the case  $m_{\gamma_*} < -1$ in the following.
    To construct a matrix $\gamma_*$ satisfying the above constraint, we write down the explicit transformation of $m$ under the 
    $\mathrm{PSL}(2,\,\mathbb{Z})$ matrix $\gamma_*\,=\,\bm{p}{q}{r}{s}$ using \eqref{frag3},
    \begin{equation}
        m_{\gamma_*} = r^2\,n + p^2\,m - p\,r\,\ell \;.
    \end{equation}~
    We will first impose the weaker condition $m_{\gamma_*} <0$ to obtain for $p/r$ (with $r \neq 0$), the range, 
    \begin{equation}\label{rcon}
        \frac{\ell}{2\,m} - \frac{\sqrt{-\Delta}}{2\,m} < \frac{p}{r} < \frac{\ell}{2\,m} + \frac{\sqrt{-\Delta}}{2\,m} \;.
    \end{equation}
       Further, under $\gamma_*$, $\ell$ transforms as 
    \begin{equation}
        \ell_{\gamma_*} =- 2\,r\,s\,n - 2\,p\,q\,m + (p\,s\,+\,q\,r)\,\ell.
    \end{equation}
    Since we only 
   count matrices in $\Gamma_+(1)$, we have
    $\ell_\gamma\,>\,0$ for all $\gamma \in W(m,n,\ell)$. This must hold, in particular, for  $\gamma_*$. Using $p s - q r = 1$, we see that this constraint can be written as
         \begin{equation}
	\frac{\ell_{\gamma_*}}{rs}  = \frac{-\Delta}{2m}   -  2m \, \left(\frac{\ell}{2m}-\frac{q}{s} \right) 
	\left(\frac{\ell}{2 m}\,-\,\frac{q}{s}\,-\,\frac{1}{rs}   \right)  >0 \;.
	  \label{lgstar}
	\end{equation}
A sufficient condition for the above inequality to hold is 
\begin{equation}
    \left(\frac{\ell}{2\,m}\,-\,\frac{q}{s} \right)\, \left(\frac{\ell}{2 m}\,-\,\frac{q}{s}\,-\,\frac{1}{r s}   \right) \leq 0,
\end{equation} 
or more explicitly,
\begin{equation}
\label{column2}
    0 \leq \frac{\ell}{2 m} - \frac{q}{s} \leq \frac{1}{rs}.
\end{equation}

    Therefore, to sum up, the condition  $m_{\gamma_*}\,<\,0$ constraints the first column $\begin{pmatrix}p\\r\end{pmatrix}$ of $\gamma_*$ according to \eqref{rcon}, while the  constraint $\ell_{\gamma_*}\,>\,0$ establishes a lower bound on its second column via \eqref{column2}. 
    A natural choice for the first column that satisfies \eqref{rcon} for arbitrary $(m, n,\ell)$ with $\Delta < 0$ 
    and saturates the second  inequality of \eqref{column2} due to  $p s -q r =1$ is
    \begin{equation}
     \begin{pmatrix}p\\r\end{pmatrix}\,=\,\begin{pmatrix} \ell/g  \\ 2m/g
     \end{pmatrix}\;,
     \label{colonestar}
    \end{equation} 
    where $g=\gcd(\ell ,2 m)$. 
    Hence our choice for the $\gamma_*$ matrix is $\bm{ \ell / g \;\; }{q}{ 2 m/ g\;\; }{s}$.
    Its conjugacy action \eqref{conjA}
on the charge bilinear matrix $A$ yields 
    \begin{eqnarray}\label{fframe}
        \gamma_*^T\,A\,\gamma_*\,=\,A_{\gamma_*}\,=\,
       \begin{pmatrix}
       2\,m_{\gamma_*} & \;\;
       -\ell_{\gamma_*}  \\
       -\ell_{\gamma_*} & \;\;
       2\,n_{\gamma_*} 
       \end{pmatrix} = 
                     \begin{pmatrix}
              2\,m\, \Delta/g^2 & \quad
               s\,\Delta/ g  \\ 
               s\,\Delta/ g & \quad
              2\,(q^2 \, m\,+\, s^2\,n - q \, s \,\ell )
              \end{pmatrix}\;.
    \end{eqnarray}
    Since $\gamma \in \Gamma_+(1)$ and 
     $\Delta <0$, the conditions $m_{\gamma_*} < 0$ and $\ell_{\gamma_* } > 0$ are indeed satisfied.
     We will now show that a choice of values of $q$ and $s$ satisfying \eqref{column2} are generated by the continued fraction representation of 
     $\ell/ 2m$.\footnote{A reader familiar with continued fractions will observe that \eqref{column2} is satisfied by the convergents \eqref{recurrel}
     of the continued fraction of $\ell/2m$.}

    \subsection{Euclid's algorithm and continued fractions}
    
  In the previous subsection, we determined a matrix   $\gamma_* =\,\bm{p}{q}{r}{s}$, whose first column is given by \eqref{colonestar}.
  The entries $q$ and $s$ of the second column have to satisfy the relation $p s - q r = 1$, but were otherwise left unspecified.
  A particular choice for $q$ and $s$ can be obtained by applying 
   Euclid's algorithm for determining $g$,  the greatest common divisor (gcd) of the two numbers $p$ and $r$ of the first column.
   An equivalent approach for determining $\gamma_*$ is provided by the continued fraction of $p/r$, as follows.
   
   Take $\ell, m \in \mathbb{N}$, and apply Euclid's algorithm to determine the gcd $g$ of the two numbers $\ell$ and $2m$.
   Euclid's algorithm can be summarized as follows:
         \begin{eqnarray}
    \ell &=& a_0 \, 2m +  r_0 \;,  \nonumber\\
    2 m&=& a_1 \, r_0 + r_1  \;,  \nonumber\\
    r_0 &=&  a_2\,r_1\,+\,r_2 \;, \nonumber\\
      r_1 &=& a_3\,r_2\,+\,r_3\;,  \nonumber\\
&\vdots& \nonumber\\
  r_{n-3} &=&  a_{n-1} \, r_{n-2} + r_{n-1} \;, \nonumber\\
      r_{n-2} &=& a_{n} \, r_{n-1} \;,
    \end{eqnarray}
with the gcd $g$ given by $g = r_{n-1}$. Note that $0 < \ell \leq m$ implies that $a_0 =0$ and $r_0 = \ell$.

The set of quotients $\{a_0, a_1,\,a_2,\, \dots, a_n\}$
is elegantly encoded in the finite continued fraction representation of $\ell/2\,m$ as 
      \begin{equation}
  \,\frac{\ell}{2m}\, = a_0 + \cfrac{1}{a_1+\cfrac{1}{a_2 + \cfrac{1}{\ddots + \cfrac{1}{a_n}} }} 
\end{equation}
which we also denote by $\ell/2m = [a_0; a_1, a_2, \dots, a_n]$. The convergents $p_k/q_k = [a_0; a_1, a_2, \dots, a_k]$
satisfy the following recursion relations \cite{halter}
 \bea
 p_{-2} &=& 0 \;\;\;,\;\;\; p_{-1} = 1 \;\;\;,\;\;\; q_{-2} = 1 \;\;\;,\;\;\; q_{-1} = 0 \;, \nonumber\\
 p_k &=& a_k \, p_{k-1} + p_{k-2} \;\;\;,\;\;\;  0 \leq k \leq n\ \;, \nonumber\\
 q_k &=& a_k \, q_{k-1} + q_{k-2}  \;\;\;,\;\;\;   0 \leq k \leq n \;,
 \label{recurrel}
 \eea
 which imply $p_0 =a_0, \, q_0 = 1$.
  In the following, we set $0 < \ell \leq m$, in which case
$\ell/2m = [0; a_1, a_2, \dots, a_n]$. 

The set $\{a_1,\,a_2,\, \dots, a_n\}$ determines the following matrix $\gamma_*$, 
  \begin{eqnarray}
        \gamma_* &=&  \begin{pmatrix}
       q  & \;\; \ell / g  \\
       s  & \;\; 2m / g 
       \end{pmatrix} = 
\bm{1}{0}{a_1}{1}\,\bm{1}{a_2}{0}{1}\,\bm{1}{0}{a_3}{1} \cdots  \bm{1}{a_n}{0}{1}
   \;\;\;,\;\;\; n \,\, {\rm even} \;,\nonumber\\
     \gamma_*  &=& 
      \begin{pmatrix}
        \ell / g  & \;\; q \\
        2m / g & \;\; s
       \end{pmatrix} = 
     \bm{1}{0}{a_1}{1}\,\bm{1}{a_2}{0}{1}\,\bm{1}{0}{a_3}{1} \cdots  \bm{1}{0}{a_n}{1}
   \;\;\;,\;\;\; n \,\, {\rm odd} \;.
   \label{decompgam}
   \end{eqnarray} 
 Observe that 
 \begin{eqnarray}
   \begin{pmatrix} \ell \\ 2m\end{pmatrix} = \gamma_* \, 
    \begin{pmatrix} 0 \\ g \end{pmatrix}   \,\, {\rm when} \, \, n \,\, {\rm even} \;,\;\;\; 
      \begin{pmatrix} \ell \\ 2m\end{pmatrix} = \gamma_* \, 
    \begin{pmatrix} g \\ 0 \end{pmatrix}   \,\, {\rm when} \, \, n \,\, {\rm odd} \;.
    \end{eqnarray}
   Note that the case $n$ even corresponds to $n_{\gamma_*} < 0$, while the case $n$ odd corresponds to $m_{\gamma_*}< 0$.
    
   The decomposition \eqref{decompgam} of $\gamma_*$ is the decomposition in the basis $U$ and $T$ as in \eqref{span}.
  Thus, Euclid's algorithm implements
successive operations of powers of $U$ and $T$. The decomposition \eqref{decompgam}
yields all the matrices
in the dyonic decay sequence $W(m, n, \ell)$ corresponding to lines of marginal stability, as in \eqref{seq}.

We now prove that $\eqref{column2}$ holds for every matrix in $W(m, n, \ell)$ given by \eqref{seq}.  Let $\gamma_i\,=\,\bm{p_i}{q_i}{r_i}{s_i}$ be the $i^{\text{th}}$ matrix in
$W(m,\,n,\,\ell)$. 
Note that $r_i >0, s_i > 0$ in view of \eqref{seq}.
Then, by construction, one  of the following two cases holds: either $\gamma_{i\,+\,1} =\gamma_{i}\,U$, in which case
\begin{eqnarray}\label{U}
p_{i\,+\,1}\,&=&\,p_i\,+\,q_i,\nonumber \\
r_{i\,+\,1}\,&=&\,r_i\,+\,s_i,\\
q_{i\,+\,1}\,&=&\,q_i,\nonumber\\
s_{i\,+\,1}\,&=&\,s_i,\nonumber
\end{eqnarray} 
or $\gamma_{i\,+\,1}=\gamma_{i}\,T$, in which case
 \begin{eqnarray}\label{T}
q_{i\,+\,1}\,&=&\,q_i\,+\,p_i,\,\nonumber \\
s_{i\,+\,1}\,&=&\,s_i\,+\,r_i,\\
p_{i\,+\,1}&=& p_i  , \nonumber\\
r_{i\,+\,i}&=& r_i.\nonumber
\end{eqnarray}

Further, let  $\gamma_{i\,+\,1}$ satisfy \eqref{column2}, implying $0 \leq \frac{\ell}{2\,m} - \frac{q_{i\,+\,1}}{s_{i\,+\,1}} \leq \frac{1}{(r_{i\,+\,1})\,(s_{i\,+\,1})}$. Then, in the case given by \eqref{U},  it follows immediately that $\gamma_i$ also satisfies \eqref{column2}, since  $r_i >0, s_i > 0$.
In the case of \eqref{T}, we see that
\begin{equation}
\frac{\ell}{2\,m} -\frac{q_i}{s_i} = \frac{\ell}{2\,m}-\frac{q_{i\,+\,1}}{s_{i\,+\,1}} - \frac{q_i}{s_i} + \frac{q_{i\,+\,1}}{s_{i\,+\,1}} \leq 
\frac{1}{r_i\,(s_{i\,+\,1})}\,+\,\frac{1}{s_i\,(s_{i\,+\,1})} = \frac{1}{r_i\,s_i} \;,
\end{equation} 
and hence $\gamma_i$ also satisfies \eqref{column2}.
Since $\gamma_*$ satisfies \eqref{column2}, the above result implies that every matrix in $W(m, n,\ell)$ does too.

Thus, we have constructed a dyonic decay sequence $W(m, n,\ell)$ of $\Gamma_+(1)$ matrices corresponding to the lines of marginal stability encountered along a continuous path from the $\mathcal{R}$-chamber to the region $*$, where the dyonic degeneracy is known or independently determined.

\subsection{Diagrammatic representation of decay walls}

The observations \eqref{U} and \eqref{T} enable us to build up a diagrammatic representation of the construction of decay walls from the continued fraction sequence.  Recall that the two columns of the matrix $\gamma=\bm{p}{q}{r}{s}$
 labelling a decay wall encode  the endpoints $q/s$ and $p/r$ of the corresponding Farey arc on the real axis in the axion-dilaton complex upper half  plane. 
 Further, the first decay wall in the sequence is $U$ corresponding to the endpoints $\frac{0}{1}$ and $\frac{1}{1}$. In the language of continued fractions, given the representation $\ell / 2m = [0; a_1, a_2, \dots, a_n]$, its $(n\,+\,1)$ convergents $p_k/q_k =[0; a_1, a_2, \dots, a_k]$ encode the decay walls, 
 with consecutive convergents $p_k/q_k$ and $p_{k-1}/q_{k-1}$ defining the
 endpoints of the decay wall (from \eqref{recurrel}, we recall the assignments $p_0/q_0 = 0/1, \, p_{-1}/q_{-1} = 1/0$).
   Consecutive convergents are related by the matrices in \eqref{decompgam} as follows,
      \begin{eqnarray}
    \begin{pmatrix}
   p_1 & p_0 \\
   q_1 & q_0
   \end{pmatrix}
   \rightarrow
     \begin{pmatrix}
   p_1 & p_2 \\
   q_1 & q_2
   \end{pmatrix}
   \rightarrow
     \begin{pmatrix}
   p_3 & p_2 \\
   q_3 & q_2
   \end{pmatrix}
   \rightarrow
     \begin{pmatrix}
   p_3 & p_4 \\
   q_3 & q_4
   \end{pmatrix} 
     \rightarrow
     \begin{pmatrix}
   p_5 & p_4 \\
   q_5 & q_4
   \end{pmatrix} \dots \;\;\;.
   \end{eqnarray}
   Hence, the continued fraction encoding of the decay walls leads to the following rule for a beautiful diagrammatic representation of $W(m, n,\ell)$ (see \cite{topology-numbers} for a detailed proof):\\

{\it The decay walls in $W(m, n,\ell)$ corresponding to consecutive convergents are edges forming a zigzag path, whose vertices are the convergents of $\frac{\ell}{2\,m}\,=\,[0;a_1,a_2,\dots,a_n]$, starting at $\frac{1}{0}$ and ending at $\frac{\ell}{2\,m}$. The path starts along the edge from $\frac{1}{0}$ to $\frac{0}{1}$, then turns left across a fan of $a_1$ triangles, then right across a fan of $a_2$ triangles etc, finally ending at $\frac{\ell}{2\,m}$.}\\

\begin{figure} [t!]
\centering
\begin{subfigure}[b]{0.4\linewidth}
\includegraphics[width=\linewidth] {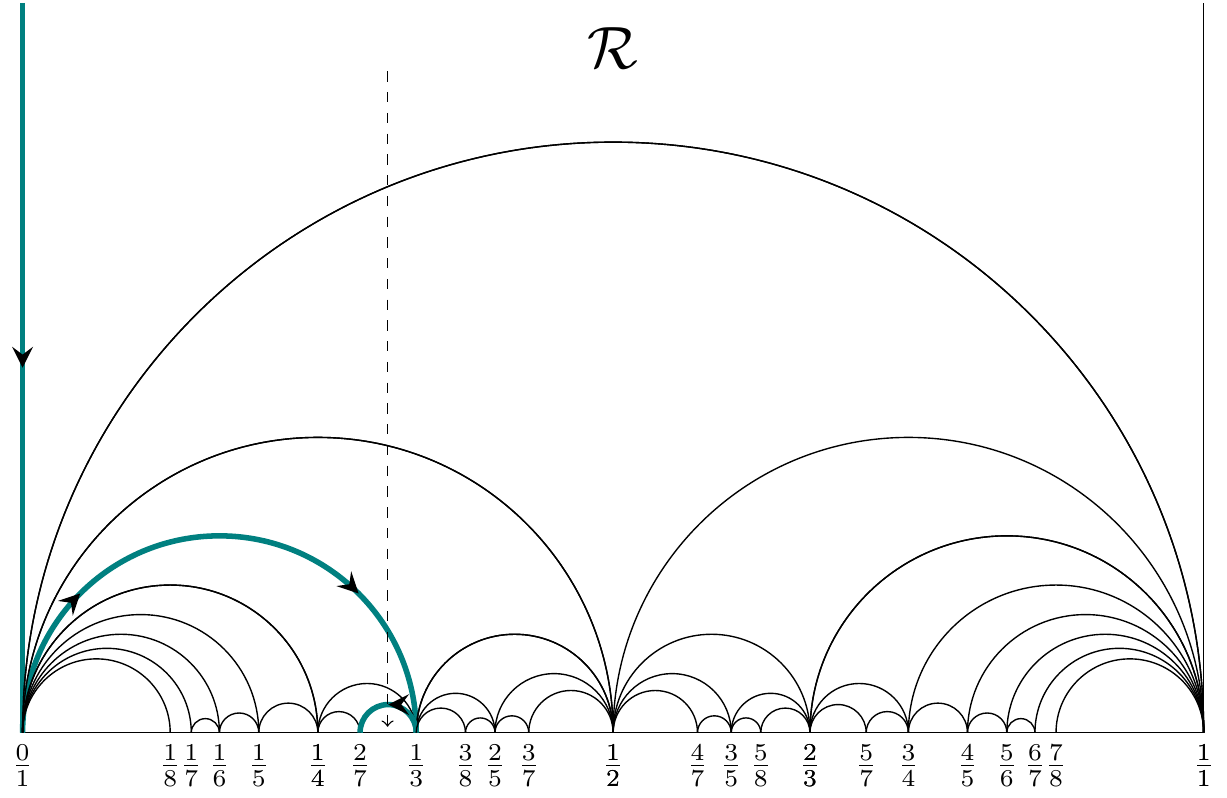}
\caption{The consecutive convergents $p_k/q_k$ and $p_{k-1}/q_{k-1}$ define the endpoints of the colored decay walls.}
\end{subfigure}
\hskip 5mm
\begin{subfigure}[b]{0.4\linewidth}
\includegraphics[width=\linewidth] {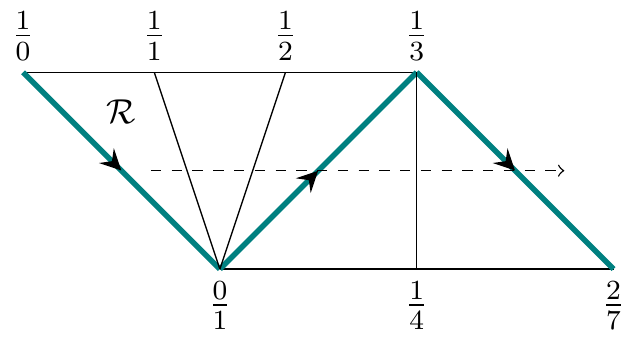}
\caption{Diagrammatic representation of the continued fraction $2/7=[0; 3, 2]$ and decay path. The number of quotients matches the number of 
	large triangles with two colored edges each. \label{tria}}
\end{subfigure}
\caption{ Characterization of the lines of marginal stability. }
\label{fig:decay}
\end{figure}

A diagram representing this rule is depicted in Figure \ref{tria} for $\ell/2m = 2/7$. Figure \ref{fig:decay} encodes an example of the complete characterization of the lines of marginal stability. Each element in $W(m, n,\ell)$ corresponds to one edge in Figure \ref{tria}. 
Hence, apart from the computation of $d_*$ in 
\eqref{truemaster}, Figure \ref{fig:decay} encodes all the information needed to solve the decadent dyon counting problem in this example.

   \subsection{Determining $d_*$  \label{sec:subD0}}

  Having characterized the dyonic decay sequence $W(m, n,\ell)$, we now turn to the determination of $d_*$ in order 
     to completely solve the dyon counting problem using \eqref{truemaster}. 
     We note that if the final charge bilinears satisfy the constraint, $m_{\gamma_*} <-1$ or $n_{\gamma_*} <-1$, then $d_* = 0$. 
   However, our construction of $W(m, n,\ell)$ only required  the matrix $\gamma_*$ 
   to satisfy the weaker condition 
   $m_{\gamma_*} 
      < 0$ or $n_{\gamma_*} < 0$.
    Therefore, we have the following two cases: 
    \begin{enumerate}
        \item 
        $m_{\gamma_*}  
               <-1$ or $n_{\gamma_*} < -1$. Hence $d_*\,=\,0$.

        \item
         $m_{\gamma_*} = 
                -1$ or $n_{\gamma_*} = -1$. Here we restrict ourselves to the case  $m_{\gamma_*} = -1$. 
       In this case, $d_* \neq 0$. In order to compute $d_*$, we consider a further action on $W(m, n,\ell)$
        by $T^j$, where $j\,>\,0$. 
               Using \eqref{frag3}, we infer the following change of the charge bilinears under the action of $T^j$, 
    \begin{equation}
        (m,\,n,\,\ell)_{\gamma_*}\,=\,(-1,\,n_*,\,\ell_*)\,\rightarrow\,(-1,\,n_*\,-j^2 - j \ell_*  = n_j,\,\ell_* + 2 j = \ell_j).
    \end{equation}
         Hence, there exists a $ j_0 >0$ such that for all $j > j_0$ we have $n_j < -1$, and hence $d(m_j =-1, n_j, \ell_j) = 0$. 
         Thus we obtain the following expression for $d_*$, 
               \begin{equation}
        d_* =   \sum_{\mu\,\in \{T,\,T^2, \dots,T^{j_0} \}} \ell_{\gamma_* \mu} \, 
       d_1(-1)\,d_1(n_{\gamma_* \mu}) = \sum^{j_0}_{j\,=\,1} (\ell_* +2 j) \, d_1(-1)\, d_1(n_* -j^2 - j \ell_*) \;.
       \label{dstexp}
    \end{equation}
    \end{enumerate}
   In the case  $n_{\gamma_*} = -1$ the above arguments go through, except that we consider the action
   on $W(m, n,\ell)$
        by $U^j$, where $j > 0$.  
 In each of the above cases, $d_*$ encodes the wall-crossing contribution to the degeneracy formula from walls corresponding to
        $\gamma_* T^j$ or $\gamma_ * U^j$ matrices. The operation by $T^j$ or $U^j$ is equivalent to extending the continued fraction of $\ell/2m$
        to  $[0; a_1, a_2, \dots, a_n, j_0]$.

          The above formula can also be extracted from $\psi_{-1} (\sigma, v)$ in \eqref{psimo}, as will be shown in section \ref{sec:disc}. 
    We have now computed all quantities on the right hand side of the dyonic degeneracy formula \eqref{truemaster} and consequently, the decadent dyon degeneracy in the $\mathcal{R}$-chamber for $\Delta < 0$. \\

      For the $\mathbb{Z}_N$ CHL orbifold models one can use the procedure of extending the continued fraction described above. 
      In this case, the wall-crossing jumps encoded by $d_*$ arise from matrices in $\Gamma_0(N)$.
             When 
       $m_{\gamma_*} <-1$ or $n_{\gamma_*} < -1/N$, 
       $d_* = 0$. If $\gamma_* \in \Gamma_0(N)$ and 
              $m_{\gamma_*} = -1$ we obtain
\begin{equation}
        d_* =   \sum_{\mu\,\in \{T,\,T^2, \dots,T^{j_0} \}} \ell_{\gamma_* \mu} \, 
       d_1(-1)\,d_2(n_{\gamma_* \mu}) = \sum^{j_0}_{j\,=\,1} (\ell_* +2 j) \, d_1(-1)\, d_2(n_* -j^2 - j \ell_*) \;.
    \end{equation}
On the other hand, if at the end of the sequence of decay walls in $W(m,n,\ell)$, corresponding to the quotients in the continued fraction of $\frac{\ell}{2m}$,  the final charge invariants satisfy  $n_* = -1/N$ with $m_*\geq -1$, we need to extend the $W$ sequence by appropriate $U$ matrices as in the $N=1$ case. In order to do so,  observe that 
the $\Gamma_0(N)$ matrix 
\begin{equation}
U^c = \begin{pmatrix}
1 & 0\\ 
c & 1
\end{pmatrix},
\end{equation}
with $c=0$ mod $N$, maps the charge bilinears $(m_*, n_*, \ell_* )$ to 
\begin{eqnarray}
m_c &=& c^2 \ n_* + m_* - c \  \ell_* \;, \nonumber\\
n_c &=& n_* \;,  \nonumber\\
\ell_c &=& \ell_* - 2 \  c \ n_* \;,
\end{eqnarray}
where $n_* = -1/N, \ m_*\geq -1, \ \ell_* \geq 0$ and $c\geq 1$.
Let $c_*$ be the minimum of the non-empty subset of natural numbers defined as $ \{ c \in \mathbb{N}, \ c = 0$ mod $ N | m_c < -1\}$. 
Then the relevant sequence of $U$ matrices to compute $d_*$ is $\{U^k|1\leq k \leq c_*, \ k = 0 $ mod $ N\}$.

   We will now display explicit formulae for decadent dyonic degeneracies with $\Delta < 0$
   for the case of heterotic string theory on $T^6$.

   \subsection{Explicit formulae \label{sec:exCHL}}

We take $0\leq \ell\leq m$ with $m >0$ as well as $\Delta < 0$. This implies $-1 \leq n < \frac14 m$. 
We denote the continued fraction of $\ell/2m$ by
\begin{equation} \label{cont-fract-explicit-formula}
	\frac{\ell}{2m} = [0;a_1,a_2,\dots,a_r].
\end{equation}
Denoting
\begin{equation}
	m_0 = m, \hspace{4mm} n_0 = n,  \hspace{4mm}  \ell_0 = \ell,
\end{equation}
we define recursively, with $1 \leq i \leq r$, 
\bea
	m_i &=& m_{i-1}+a_i^2n_{i-1}-a_i\ell_{i-1}, \hspace{4mm} n_i = n_{i-1}, \hspace{4mm} \ell_{i} = \ell_{i-1}-2a_in_{i-1}, \hspace{4mm}\text{ for $i$ odd}
	\nonumber\\
	m_i &=& m_{i-1}, \hspace{4mm} n_i =a_i^2m_{i-1}+ n_{i-1}-a_i\ell_{i-1}, \hspace{4mm} \ell_{i} = \ell_{i-1}-2a_im_{i-1}, \hspace{4mm}\text{ for $i$ even}.
	\nonumber\\
\eea
Note that this is the same as computing the matrices $\gamma_i$ (using \eqref{span})  from the continued fraction \eqref{cont-fract-explicit-formula} and defining
\begin{equation}
	m_i = m_{\gamma_i}, \hspace{4mm} n_i = n_{\gamma_i}, \hspace{4mm} \ell_i = \ell_{\gamma_i}.
\end{equation}
Lastly, define
\bea
	m_{ij} &=& m_{i-1}+j^2n_{i-1}-j\ell_{i-1}, \hspace{2mm} n_{ij} = n_{i-1}, \hspace{2mm} \ell_{ij} = \ell_{i-1}-2jn_{i-1}, \hspace{2mm}\text{ for $i$ odd and } 1\leq j \leq a_i, \nonumber\\
	m_{ij} &=& m_{i-1}, \hspace{2mm} n_i =j^2m_{i-1}+ n_{i-1}-j\ell_{i-1}, \hspace{2mm} \ell_{ij} = \ell_{i-1}-2jm_{i-1}, \hspace{2mm}\text{ for $i$ even and } 1\leq j \leq a_i. \nonumber\\
\eea
Then we obtain
for the decadent dyonic degeneracy \eqref{truemaster},
\begin{equation}
	d(m,n,\ell) = d_* +  (-1)^{\ell +1} \, \sum_{i=1}^r\sum_{j = 1}^{a_i} | \ell_{ij} | \, d_1(m_{ij}) d_1(n_{ij}) \;.
\end{equation}
Here the summation is only over values $m_{ij}$ and $n_{ij}$ satisfying $m_{ij}, n_{ij} \geq -1$. 

In the very specific instance when
$m_{\gamma_*} = -1$, we see that $n = \tfrac14 (m-1)$ and $\ell = m$. Hence $m+1$ is even.
The continued fraction of $\ell/2m$ is then simply $\ell/2m = [0;2]$, and the decadent dyonic degeneracy \eqref{truemaster} takes the form
\begin{equation}
d(m,n,\ell) = \Big(  \sum_{q=1}^{  \left \lfloor {  \sqrt{ \frac{m}{4} + 1} - \tfrac12 }\right \rfloor 
	} (2 q + 1) \, d_1( n - q^2 - q) \Big)  + \tfrac12 (m+1) \, \left( d_1(n) \right)^2 + d_1(n)
	\;.
	\end{equation}

\subsection{ $\Delta = 0$ \label{sec:del0}}

Now we consider the case $\Delta = 4 m n - \ell^2=0$. We restrict to $0 \leq \ell \leq m$ with $m> 0$, without loss of generality, following \cite{Chowdhury:2019mnb}.

   The logic and computational steps are exactly as in the $\Delta < 0$ case. The $\gamma_*$ frame charge bilinear matrix \eqref{fframe} takes on the value
    \begin{equation}\label{fff}
     A_{\gamma_*} = \bm{0}{0}{0}{\; 2\,n_{\gamma_*}} \;\; \lor \;\;   A_{\gamma_*} = \bm{2m_{\gamma_*} \; }{0}{0}{0} \;.
    \end{equation}
    Notice that the conjugacy operation in $\eqref{fframe}$ preserves not just the determinant $\Delta$ of the bilinear matrix, but also the greatest common divisor $\tilde{g}$ of its elements. 
   Denoting $\gcd(m, n,\ell ) = \Tilde{g}$, we see that the only non-zero entry of $A_{\gamma_*}$ in \eqref{fff} is 
    \begin{equation}
 2\,\tilde{g} = 2\,n_{\gamma_*} \;\; \lor \;\;  2\,\tilde{g} = 2\,m_{\gamma_*} \;.
    \end{equation}
As before, the continued fraction representation of $\ell / 2 m = [0; a_1, \dots, a_n]$
yields a sequence of convergents $p_k/q_k = [0; a_1, \dots, a_k], \, k \leq n$,
corresponding to the dyonic decay walls, with the last wall $\gamma_*$  representing an immortal $1/4$ BPS dyon
$(0, \Tilde{g}, 0)$ or $({\tilde g}, 0, 0)$. Notice that  $(0, \Tilde{g}, 0)$ can be transformed into $({\tilde g}, 0, 0)$ by the matrix
 $S \in \mathrm{PSL}(2, \mathbb{Z})$, which is a symmetry of heterotic string theory on $T^6$. Hence $d (0,\Tilde{g}, 0)  = d ({\tilde g}, 0, 0)$, and
 without loss of generality, we will now consider the case $(0,\Tilde{g}, 0)$.

Using the  expansion of the dyonic degeneracy generating function for heterotic string theory on $T^6$
in powers of $m$ \cite{Dabholkar:2012nd}\footnote{See section \ref{sec:disc}  for an expanded discussion.},
\begin{eqnarray}
\frac{1}{\Phi_{10}(\rho, \sigma, v)} &=&  \psi_{-1} e^{- 2 \pi i  \rho} +  \sum_{m\,=\,0}^{\infty}  \left(\psi^F_{m}(\sigma,\,v) + \psi^P_{m}(\sigma,\,v)
 \right) e^{2 \pi i m \rho} \;, \nonumber\\
\psi^F_{0}(\sigma) &=& 2\,\frac{E_2(\sigma)}{\eta^{24}(\sigma)},
\end{eqnarray}
it can be seen 
that  
the degeneracy of these $m_{\gamma_*} = 0,\,\,\ell_{\gamma_*} = 0$  dyons in the $\mathcal{R}$-chamber is captured by the quasi modular form, 
\begin{equation} 
2\,\frac{E_2(\sigma)}{\eta^{24}(\sigma)} = -\, 2\,\sum_{n\,\geq\,-1} n\,d_1(n) \, q^n \;.
\label{d1e2}
\end{equation}
Hence, $d_* = (-1)^{\ell} \, 2\,\Tilde{g}\,d_1(\Tilde{g})$.
 The dyonic degeneracy formula for zero discriminant states in the $\mathcal{R}$-chamber is 
\begin{equation}\label{d0}
d(m, n, \ell) = 2\,\Tilde{g}\,d_1(\Tilde{g}) 
- \sum_{\gamma\,\in \,W(m, n, \ell)} \, | \ell_{\gamma} |  \, d_1(m_\gamma)\,d_1(n_\gamma) 
 \;.
\end{equation}
Here we used the fact that for $\Delta =0$, $\ell = 0 \mod 2$. This implies $d_* > 0$, as is the case for single centre $1/4$ BPS black 
holes \cite{Sen:2010mz}.

This solves the exact decadent dyon counting problem for the $1/4$ BPS states characterized by $I=1$ in heterotic string theory on $T^6$.

\subsection{Decadent dyons with torsion $I > 1$ \label{sec:decd}}

The extension of our solution to the dyon counting problem for $I \neq 1$ in heterotic string theory on $T^6$
is quite  straightforward. We again restrict to $0 \leq \ell \leq m$ with $ m>0$.
The dyonic degeneracy, $d_I (m, n, \ell)$ for general $I$ is expressed in terms \cite{Banerjee:2008pu,Dabholkar:2008zy} of the dyonic degeneracy of states $d(m, n, \ell)$ for $I =1$  as
\begin{equation}\label{I}
    d_I(m,n,\ell)\,=\, \sum_{t|I}\,  t \, d(m, \frac{n}{t^2}, \frac{\ell}{t}) \;.
\end{equation}
Each summand in \eqref{I}  is computed by \eqref{truemaster} via the continued fraction approach to generating decay walls. 
However, note that only those decay walls contribute to \eqref{truemaster} that correspond to $n_\gamma,\, \ell_\gamma\,t = 0 \mod  t^2$. This condition implies that the dyonic walls correspond to a subset of the $\mathrm{PSL}(2,\,\mathbb{Z})$ group, which as we show in the appendix, is the subgroup 
\begin{equation}
    \Gamma^0(t)\,=\,\left\{\bm{a}{b}{c}{d}\,\in \,\mathrm{PSL}(2,\,\mathbb{Z})\,|\,b =0 \mod t \right\}.
\end{equation} 
Hence, choosing only $\Gamma^0(t)$ matrices from the continued fraction algorithm, we can perform an exact computation of $d_I$, using \eqref{I}. 
We note here that each $t | I $ summand in \eqref{I}   is computed via the continued fraction representation of $\ell / (2\,t\,m)$
(see \eqref{zt}), while the constraint \eqref{column2} is implemented as $0 \leq \frac{\ell}{2\, t \, m } - \frac{q}{s}\leq\,\frac{1}{r\,s}$. Further, the terminal point $*$ in this case corresponds to $m_{\gamma_*} < -1$ or $n_{\gamma_*} < -t^2$. It can be shown that
 $m_{\gamma_*} < -1$ for the sectors $t >1$. This implies that the contribution from the sectors $t >1$ is entirely determined by the set of matrices
 generated by the continued fraction of $\ell / (2\,t\,m)$, while the sector $t =1$ requires computing $d_*$ as in \eqref{dstexp}.

We have thus solved the decadent dyon counting problem for all negative and zero discriminant $1/4$ BPS states in $\mathcal{N} = 4$ 
heterotic string theory on $T^6$.  We now turn to its CHL orbifolds for $N = 2,\,3,\,5,\,7$.

\subsection{$\mathbb{Z}_N$ CHL orbifold models  \label{sec:CHL}}

\begin{figure} [t!]
\centering
\begin{subfigure}[b]{0.6\linewidth}
\includegraphics[width=\linewidth] {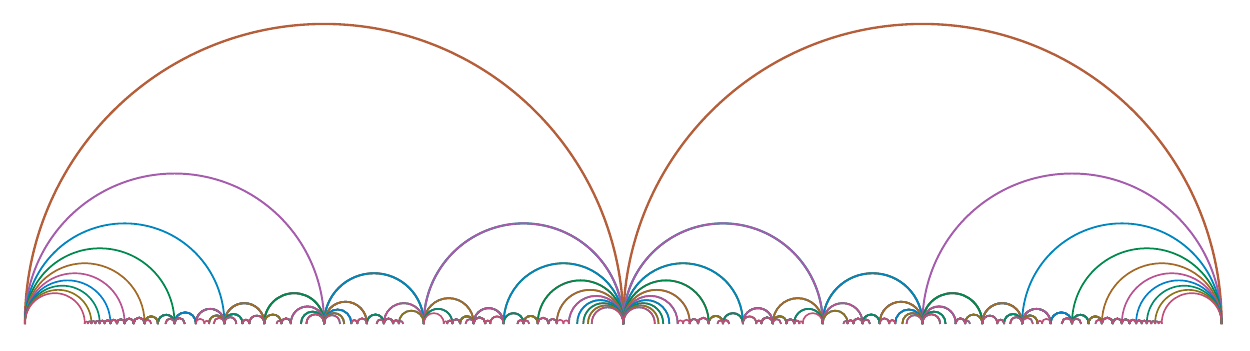}
\end{subfigure}
\hskip 5mm
\begin{subfigure}[b]{0.6\linewidth}
\includegraphics[width=\linewidth] {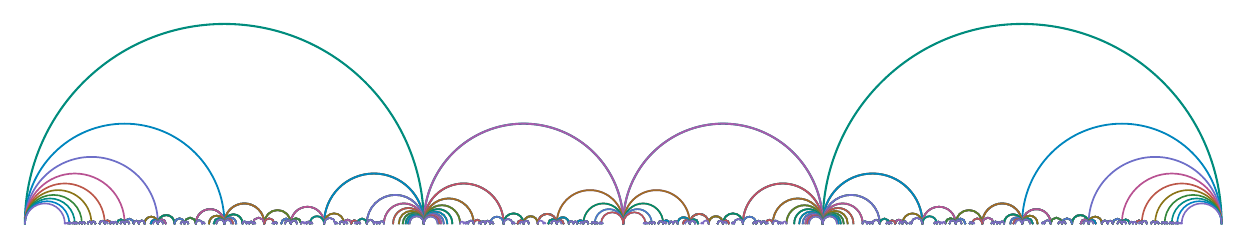}
\end{subfigure}
\hskip 5mm
\begin{subfigure}[b]{0.6\linewidth}
\includegraphics[width=\linewidth] {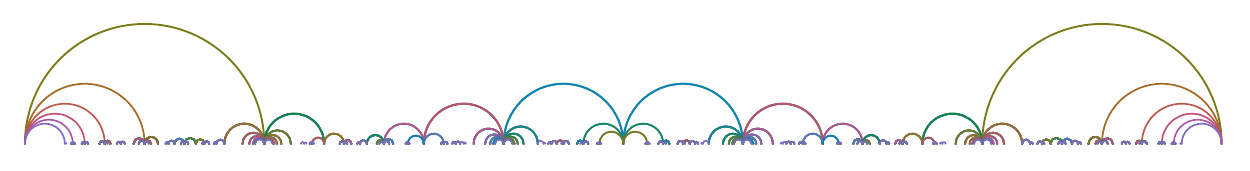}
\end{subfigure}
\hskip 5mm
\begin{subfigure}[b]{0.6\linewidth}
\includegraphics[width=\linewidth] {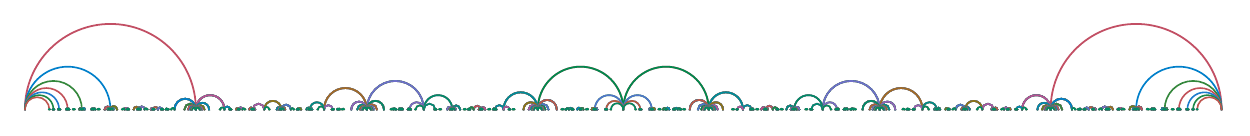}
\end{subfigure}
\caption{Lines of marginal stability in the $R$-strip in $\mathbb{Z}_N$ CHL models with
$N = 2, 3, 5, 7$ from top to bottom.
 }
\label{fig:wallsN}
\end{figure}

We consider $\mathbb{Z}_N$ CHL orbifold models with $N =2,3,5, 7$.
The lines of marginal stability in these models have been extensively analyzed in \cite{Sen:2007vb,Sen_2011}, 
where it was shown that the corresponding matrices lie in the congruence 
subgroup $\Gamma_0(N) \subset \mathrm{PSL}(2,\,\mathbb{Z}) $ defined in \eqref{gam0}.\footnote{Unlike in the parent heterotic theory, this does not coincide with the electric-magnetic duality group 
$\Gamma_1(N) =\{\bm{a}{b}{c}{d} \in 
\mathrm{PSL}(2,\,\mathbb{Z})  | c= 0 \mod N, \;\; a, d = 1 \mod N \}$ and is an 'accidental' symmetry of the exact degeneracy formula.}
Some of the lines of marginal stability in these models are shown in Figure \ref{fig:wallsN}.

We compute the degeneracies for twisted sector torsion $I=1$ decadent dyons in the $\mathcal{R}$-chamber with
$0 \leq \ell <  2 m$. As in the heterotic string theory on $T^6$, we consider the 
continued fraction of $\ell/2m$.

For our purposes, we need only observe that unlike in the parent heterotic theory, the $\mathrm{PSL}(2,\,\mathbb{Z})$ $S$-matrix is no longer a symmetry
when $N > 1$. Consequently, given a wall $\gamma$ labelled by a $\Gamma_+(N)$ matrix, the $S$-dual wall 
 $\gamma S$ need not lie in $\Gamma_-(N)$, and hence is no longer guaranteed to be a legitimate decay wall.
 This implies that we must independently count contributions from the $\Gamma_+(N)$ walls with 
$\ell_\gamma > 0$ and $\Gamma_-(N)$ walls with $\ell_\gamma < 0$ in the decay sequence 
${W}_N(m, n, \ell)$. We further note that the continued fraction representation of $\ell/2m$ generates non-negative quotients, and hence generates walls in $\Gamma_+(1)$ with $\ell_\gamma > 0$, by construction. In the orbifold theories, the $\Gamma_-(N)$ contributions can be extracted as follows. Let
\begin{equation}
    \Omega_N\,=\,\left\{\gamma\,\in\, \Gamma_+(1) = \bm{a}{b}{c}{d}  \in \, \mathrm{PSL}(2,\,\mathbb{Z})
    | \, d = 0 \mod N \right\}.
\end{equation}
Then, given a $\Omega_N$ matrix, $\gamma = \bm{a}{b}{c}{d}$, labelling a decay wall generated by the continued fraction representation of 
$\frac{\ell}{2\,m}$ 
and corresponding to $\ell_\gamma > 0$, we conclude that it will not represent a legitimate dyonic decay wall for the $\mathbb{Z}_N$ CHL orbifold. However, the $S$-dual matrix $\gamma\,S$ lies in $\Gamma_-(N)$ with $\ell_\gamma < 0$ and will constitute a valid wall. Hence, we can sum up the decadent dyon counting solution for CHL models as follows:
\newline
\newline{\it The dyonic decay sequence $W_N(m,n,\ell)$ in a $\mathbb{Z}_N$ CHL orbifold of the heterotic string theory on $T^6$ is given by
\begin{equation}
   W_N (m, n, \ell) = \left( W\,\cap\,\Gamma_+(N) \right)  \cup \left(  W S \,\cap\,\Gamma_-(N) \right) \;,
   \label{setWN}
\end{equation}}
where $W$ denotes the set $W(m, n,\ell)$ in heterotic string theory on $T^6$.

Having defined $W_N$, decadent dyonic degeneracies can be computed for $I =1$ dyons in these models by \eqref{truemaster}. 
We note that for  $\Delta =0$ dyons, the degeneracy $d_*$ of a dyon with charge bilinears $(0, {\hat g}, 0)$,
where
\bea
{\hat g} = \frac{\gcd(m, n N, \ell)}{N} \;,
\eea
is obtained as follows. Similar to the $N=1$ case, 
the expansion \eqref{phikpsi} of 
the modular form $1 / \Phi_k$, given in \eqref{1},  that counts $1/4$ BPS dyons 
yields the following $m=0$ quasi-modular counting function\footnote{ $\psi^F_{-k,\,0} $ in \eqref{psik0N} is the Fricke dual of (A.73) in 
\cite{Bossard:2018rlt}.} 
\cite{Bossard:2018rlt},  
\begin{equation}
    \psi^F_{-k,\,0}(\sigma)\,=\, \frac{k+2}{12\,(N\,-\,1)}\,\frac{E_2(\sigma/N ) -E_2( \sigma)}{\eta^{k+2}(\sigma/N)\,
    \eta^{k+2}(\sigma)} 
    = \sum_{n N \in \mathbb{N}_0 }\, d_N(n)\,q^n \;,
    \label{psik0N}
\end{equation}
where we recall the relation $24/(N+1) = k + 2$.
Hence, $d_* = (-1)^{\ell +1} d_N ({\hat g}) $. Using 
\eqref{truemaster},
the $\Delta =0$ CHL decadent dyonic degeneracy is then given by 
\begin{equation}
    d(m,\,n,\,\ell) = - \left( d_N ({\hat g}) +
     \sum_{\gamma\,\in\,W_N(m,n,\ell)} \, | \ell_{\gamma} | \, d_1(m_\gamma)\,d_2(n_\gamma)\
   \right) \;,
\end{equation}
where we used that for $\Delta =0$, $\ell = 0 \mod 2$.

We have performed extensive numerical checks on the correctness of \eqref{truemaster} in $\mathbb{Z}_N$ CHL orbifold models 
with $N=1, 2, 3, 5, 7$.

\subsection{Examples}

We illustrate below the above procedure of constructing $W(m, n, \ell)$ respectively $W_2(m, n, \ell)$ to count decadent dyonic degeneracy by an example each of the $\Delta<0$ and $\Delta=0$ cases in heterotic string theory on $T^6$ as well as an example in the $\mathbb{Z}_2$ CHL orbifold.
\begin{enumerate}
    \item Heterotic on $T^6$:
\begin{enumerate}
\item $(m,n,\ell) = (14,1,8)$ \\
Then, $\Delta = -8$ and 
$\frac{\ell}{2m} = \frac{2}{7} = [0;3,2]$
yielding walls corresponding to 
\begin{equation}
	\begin{pmatrix}
		1 & 0\\
		1 & 1
	\end{pmatrix}, \begin{pmatrix}
		1 & 0\\
		2 & 1
	\end{pmatrix},
	\begin{pmatrix}
		1 & 0\\
		3 & 1
	\end{pmatrix},
	\begin{pmatrix}
		1 & 0\\
		3 & 1
	\end{pmatrix}\begin{pmatrix}
		1 & 1\\
		0 & 1
	\end{pmatrix} = 
	\begin{pmatrix}
		1 & 1 \\
		3 & 4
	\end{pmatrix},
	\begin{pmatrix}
		1 & 0\\
		3 & 1
	\end{pmatrix}\begin{pmatrix}
		1 & 2\\
		0 & 1
	\end{pmatrix} =
	\begin{pmatrix}
		1 & 2\\
		3 & 7
	\end{pmatrix}
	\label{matex}
\end{equation}
and respective charge bilinears
\begin{equation}
	(7,1,6), (2,1,4), (-1,1,2), (-1,-2,4), (-1,-7,6)\,.
\end{equation}
The dyonic degeneracy is hence computed to be 
\begin{equation}
	d(14,1,8) =(-1)\left( 6 d_1(7)d_1(1)+4 d_1(2)d_1(1)+2d_1(-1)d_1(1)\right) = -58\,671\,297\,648.
\end{equation}

\item $(m,n,\ell) = (49,4,28)$.\\
Then, $\Delta = 0$ and $\frac{\ell}{2m} = \frac{2}{7} = [0;3,2]$, which yields the same walls as in \eqref{matex}.
The respective charge bilinears are now
\begin{equation}
	(25,4,20), (9,4,12), (1,4,4), (1,1,2), (1,0,0).
\end{equation}
Further, 
\begin{equation}
    d_*(49,4,28) = d(1,0,0) = 2 d_1(1)= 648 \,.
\end{equation}
This yields the dyonic degeneracy to be 
\begin{eqnarray}
	d(49,4,28) &=& 648-\left(20 d_1(25)d_1(4)+12 d_1(9) d_1(4)+ 4 d_1(1) d_1(4)+ 2 d_1(1) d_1(1)  \right) \nonumber\\
	&= & -459\,542\,242\,945\,399\,203\,613\,080.
\end{eqnarray}

\item $(m,n,\ell) = (12,3,12)$.\\
Then, $\Delta = 0$ and $\frac{\ell}{2m} = \frac{1}{2} = [0;2]$ yielding walls,
\begin{equation}
	\begin{pmatrix}
		1 & 0\\
		1 & 1
	\end{pmatrix}, \begin{pmatrix}
		1 & 0\\
		2 & 1
	\end{pmatrix}
\end{equation}
and respective charge bilinears,
\begin{equation}
	(3,3,6), (0,3,0).
\end{equation}
Then, 
\begin{equation}
 d_*(12,3,12) = d(0,3,0) = 6 d_1(3)= 153\,900  \,,
\end{equation} 
so that the dyonic degeneracy is 
\begin{equation}
	d(12,3,12) = 153\,900 -6 d_1(3)d_1(3) = -3\,947\,381\,100\,.
\end{equation}
In all the above cases, the computed decadent dyonic degeneracy tallies with that read off from the corresponding Fourier coefficient of $\frac{1}{\Phi_{10}}$.
\end{enumerate}

\item $\mathbb{Z}_2$ CHL orbifold:

$N=2$:
$(m,n,\ell) = (7,\frac{1}{2},4)$.\\
Then, $\Delta = -2$ and $\frac{\ell}{2m} = \frac{2}{7} = [0;3,2]$, which yields the same matrices as in \eqref{matex}.
Out of these, the matrices that are relevant for the degeneracy computation are those that lie in 
$W_2=( W\cap\Gamma_+(2) ) \cup ( WS\cap\Gamma_-(2))$, namely,
\begin{equation}
	 \begin{pmatrix}
		1 & 0\\
		2 & 1
	\end{pmatrix},
	\begin{pmatrix}
		-1 & 1 \\
		-4 & 3
	\end{pmatrix}
\end{equation}
with respective charge bilinears
\begin{equation}
	(1,\frac{1}{2},2), (-1,-\frac{1}{2},-2).
\end{equation}
This yields, 
\begin{equation}
	d(7,\frac{1}{2},4) = -2d_1(1)d_2\left(\frac{1}{2}\right)-2d_1(-1)d_2\left(-\frac{1}{2}\right) = -5410 \,.
\end{equation}
 The computed decadent dyonic degeneracy tallies with that read off from the corresponding Fourier coefficient of $\frac{1}{\Phi_{6}}$.
 
 \end{enumerate}

In the above, although we have, for reasons of calculational simplicity, demonstrated examples with small values of the charge invariants, it is worth pointing out that the algorithm developed here is, in fact, a computationally more efficient way to compute degeneracies of negative discriminant states than a brute force expansion of the inverse of the Siegel form, $1/\Phi_k$,
especially for large charge invariants.  An illustrative example is provided by a class of negative discriminant states with $\ell = m$ and $n<m/4$.  For large values of $\ell$ and $m$, extracting the corresponding coefficient of 
$1/\Phi_k$ is indeed a time-consuming operation. However, in our algorithm, we observe that the continued fraction of the ratio $\frac{\ell}{2m}= \frac{1}{2}$ is simply $[0; 2]$, which generates at most two lines of marginal stability represented by the matrices $U$ and $U^2$. In particular, as $m_{U^2}=\Delta/m<-1$,  there is only one line of marginal stability that contributes to the degeneracy for $N=1$, while for $N>1$ there will be no contribution, since $U\notin \Gamma_0(N)$ for $N>1$, and hence the degeneracy will be zero. 
This allows us to trivially compute the corresponding charge bilinears as $(m_U = n, n_U = n, \ell_U = m-2n)$. All that is left in order to compute the degeneracy jump at the $U$ decay wall is to simply obtain the coefficient of $q^{n}$ in the Fourier expansion of the modular form $1/\eta^{24}(\rho)$ (c.f. \eqref{Pffk}),  a far more temporally efficient strategy than a brute force series expansion of the inverse of the Siegel form in three variables.

\section{Discussion \label{sec:disc}}

We systematize various features of the reasoning behind and implications of our results for  heterotic $\mathbb{Z}_N$ CHL models below.

The microstate degeneracies $d(Q,P)$ of 
$\frac14$ BPS 
dyons (with $I = 1$) in heterotic string theory on $T^6$
 are determined in terms of three charge bilinears, denoted by $m, n, \ell$ in subsection \ref{sec:notset}, i.e. $d(Q, P) = d(m, n, \ell)$. 
 These bilinears 
satisfy the bounds
$m \geq -1, n \geq -1$.  
$\frac14$ BPS dyonic states can be classified into immortal and decadent states. Immortal states exist at all points in the axion-dilaton moduli space. These are either single centre $\frac14$ BPS black holes  which require $m, n > 0$ as well as $\Delta = 4 m n - \ell^2 >0$ (positive discriminant states) or zero discriminant states with $\Delta = 0$.
 
Decadent states are two-centred $\frac14$ bound states of $\frac12$ BPS constitutents that either cease to exist or come into existence when crossing walls of marginal stability in the axion-dilaton moduli space \cite{Sen:2007vb,Sen_2011}.
The generating function $1/\Phi_{10}$ of $\frac14$ BPS dyonic degeneracies for $I\,=\,1$ in heterotic string theory on $T^6$
is defined
on the Siegel upper half plane of genus $2$ and admits the Fourier expansion 
\be
\frac{1}{\Phi_{10} (\rho, \sigma, v)} = \sum_{m=-1}^{\infty} \psi_m (\sigma, v) \, e^{2 \pi i m \rho} \;,
\label{fexp10}
\ee
where the Fourier-Jacobi coefficients $\psi_m$ are meromorphic Jacobi forms of weight $-10$ and index $m$ with respect to $\mathrm{PSL}(2, \mathbb{Z})$.
For evaluating their microstate degeneracies, the symmetries of the even weight Jacobi forms $\psi_m$ enable us to choose $\ell$ to lie in the range $0\, \leq\, \ell\,<\, 2m$, and subsequently
restrict $\ell$ to lie in the range $0 \leq \ell \leq m$, with no loss of generality \cite{Chowdhury:2019mnb}.
As shown in \cite{Dabholkar:2012nd}, the $\psi_m$ with $m\geq 0$ have a canonical decomposition  into two mock Jacobi forms,
\begin{equation}\label{split mod}
    \psi_m\,=\,\psi_m^F\,+\,\psi_m^P,
\end{equation}
with $\psi_m^F$ and $\psi_m^P$,
referred to as the finite and polar parts respectively. 
The finite part $\psi_m^F$, being holomorphic, possesses the
Fourier expansion 
\be\label{mfourie}
\psi_m^F (\sigma, v) = \sum_{n, \ell \in \mathbb{Z}} \, c_m^F (n,\ell) \, q^n \, y^{\ell}  \;\;\;,\;\;\; 
 q = e^{ 2 \pi i \sigma} \;\;\,,\;\;\; y= e^{2 \pi i v} \;,
\ee
with well-defined Fourier coefficients $c_m^F (n,\ell)$. 
We computed the dyonic degeneracy in a region of the moduli space called $\mathcal{R}$-chamber, where the decadent contribution to dyonic degeneracy arises from two kinds of two-centred states, namely $\frac14$ BPS states with $\Delta < 0$ 
and $\Delta=0$.

Thereby, we showed that the contribution of negative discriminant states to the index $d(m, n, \ell)$ is 
determined in terms of the continued fraction of the rational number $\ell/2m$ and in terms of 
T-walls associated with the Fourier coefficients $c_{-1} (n,\ell)$ of $\psi_{-1}$. 
The microstate degeneracies of single centre $\frac14$ BPS black holes are encoded in the  mock modular forms $\psi_m^F$ \cite{Dabholkar:2012nd}.
These degeneracies turn out to be determined in terms of negative discriminant states  \cite{Ferrari:2017msn,Chowdhury:2019mnb}
through a generalized Rademacher expansion of these mock modular forms \cite{bringono,bringmann2010sheaves,bringono2,bringmahl}.
Hence, our result can be parsed as: \\

{\it Single centre $\frac14$ BPS black hole degeneracies with $I=1$
are determined in terms 
of the continued fraction of the rational number $\ell/2m$ and 
walls corresponding to $T$ and $U$ matrices (c.f. \eqref{seq}) associated with the Fourier coefficients $c_{-1} (n,\ell)$ of $\psi_{-1}$}.\\

We proceed to summarise the chain of steps that establish this remarkable result.  In what follows, we will first identify the ${R}$-strip as one of the strips into which the axion-dilaton moduli space is divided into by lines of marginal stability generated by powers of $T$ (see \eqref{Twallk})
and derive a formula for $c_{-1}$ in terms of wall-crossing across these lines. 
The Fourier coefficients of $\psi_m^F$ in \eqref{mfourie} can be divided into 
3 sets which are characterized by
whether $\Delta \equiv 4 m n -\ell^2$ is positive, zero 
or negative.  
 The polar part $\psi_m^P$ that appears in the decomposition of $\psi_m$ (with $m \geq 0$) is given in terms of an Appel-Lerch sum ${\cal A}_{2,m}$ \cite{Dabholkar:2012nd},
 \be
 \psi_m^P (\sigma, v) = \frac{d(m)}{\eta^{24} (\sigma)} \, {\cal A}_{2,m} (\sigma, v) \;,\;
 {\cal A}_{2,m} (\sigma, v) =  \sum_{s \in \mathbb{Z}} \frac{ q^{m s^2 + s} \, y^{2 m s + 1}}{ (1 - q^s \, y)^2} \;,\;
 q = e^{ 2 \pi i \sigma} \;,\; y= e^{2 \pi i v} \;,
 \label{psipol}
 \ee
 where\footnote{Here $d(m) = d_1(m)$, with $d_1(m)$ given by \eqref{d1e2}.} $d(m)$
  denote the Fourier coefficients of $1/\eta^{24} (\sigma)$, i.e. $ 1/\eta^{24} (\sigma) = \sum_{m \geq -1} d(m) q^m$.
 The Appel-Lerch sum ${\cal A}_{2,m}$ exhibits wall crossing, which means that its Fourier coefficients are only well defined in a range 
 $s < {\rm Im } v/{\rm Im} \sigma < s+1, 
 \, s \in \mathbb{Z}$, in which case 
 \be
\psi_m^P (\sigma, v) = \sum_{n, \ell \in \mathbb{Z}} \, c_m^P (n,\ell) \, q^n\, y^{\ell} \;.
\ee
For fixed $s$, this condition on the range of $ {\rm Im } v/{\rm Im} \sigma $
defines a strip in the upper-half complex plane $\Sigma = \Sigma_1 + i \Sigma_2$ (also called axion-dilaton moduli space), with 
 $\Sigma_1 = - {\rm Im} v / {\rm Im} \sigma$.
 The boundaries of these strips are referred to as T-walls, and the Fourier expansion coefficients
 of  ${\cal A}_{2,m}$ suffer jumps when one crosses a T-wall. These jumps are interpreted \cite{Dabholkar:2012nd}
 as being due to the existence of a bound state
 of two $\frac12$ BPS black holes on one side of the T-wall that ceases to exist when crossing the wall to the other side.
The strip $s=-1$ is referred to as the ${R}$-strip. The $\mathcal{R}$-chamber is the region in the ${R}$-strip adjoining
the elementary T-wall defined by $\Sigma_1 =0$ and which does not contain lines of marginal stability.

 The Fourier-Jacobi coefficient $\psi_{-1}$ appearing in \eqref{fexp10} is special in that its decomposition does not contain a finite part. 
 $\psi_{-1}$ is given by $\psi_{-1} (\sigma, v) = \phi_2 (\sigma, g)/ \eta^{18} (\sigma)$, where $\phi_2$ denotes
 the Jacobi form of index $-1$ given by
$\phi_2 = 1/\vartheta_1^2(\sigma,v)$, which  possesses the following
 representation in terms of Appel-Lerch type sums \cite{Bringmann:2014nba} 
 (we use the definition of the Jacobi theta function $\vartheta_1$ given in \cite{Jatkar:2005bh}),
 \be
 \phi_2 (\sigma, v) = \eta^{-6} (\sigma) \, 
 \sum_{n \in \mathbb{Z}} (2n + 1) \frac{q^{n (n+1)} y}{1 -  q^n y} +  \sum_{n \in \mathbb{Z}}  \frac{q^{n^2 + 2n} y^2}{(1 -  q^n y)^2}
 \;\;\;,\;\;\; 
 q = e^{ 2 \pi i \sigma} \;\;\,,\;\;\; y= e^{2 \pi i v} \;.
 \label{phi2exp}
 \ee
 
Following \cite{Sen_2011,Chowdhury:2019mnb}
we chose to work in the $\mathcal{R}$-chamber.
Evaluating the Appel Lerch sum ${\cal A}_{2,m}$ (with $m \geq 0$) in this chamber yields
\be
 {\cal A}_{2,m} (\sigma, v) = \left( \sum_{k >0} \sum_{l>0} - \sum_{k \leq 0} \sum_{l<0} \right) \, l \, q^{m k^2 + l k} \, y^{2 m k +l} \;.
  \ee
Setting $\ell = l + 2 m k$, one readily observes that there are no terms $y^{\ell}$ in this expansion with 
$\ell$ in the range $0 \leq \ell < 2m$. Hence, the Fourier coefficients $c_m^P (n,\ell)$ of 
$ \psi_m^P (\tau, z) $ vanish for $\ell$ in the range $0 \leq \ell < 2m$
when evaluated in the $\mathcal{R}$-chamber.
Consequently, 
for this range of values of $\ell$, the Fourier coefficients of $\psi_m$ 
in the $\mathcal{R}$-chamber
 equal the Fourier coefficients
$c_m^F (n,\ell)$ of $\psi_m^F$  \cite{Chowdhury:2019mnb}. Therefore, there are no contributions from $\psi_m^P$ to 
decadent dyonic degeneracies in the $\mathcal{R}$-chamber.

In the $\mathcal{R}$-chamber,
$ \eta^6 (\sigma) \, \phi_2 (\sigma, v) $ has the Fourier expansion
\begin{equation}
 \eta^6 (\sigma) \, \phi_2 (\sigma, v) =	\sum_{\ell\geq 1} \ell\, y^{-\ell} + \sum_{k\geq 1} 2k \,q^{k^2} + \sum_{\ell\geq 1}\sum_{k\geq 1}(2k+\ell)
  q^{k^2+\ell k}(y^\ell + y^{-\ell}), \;
  \label{etphexp}
\end{equation}
while 
the Fourier expansion of $\psi_{-1}$,
\be
\psi_{-1} (\sigma,v) =  \sum_{n, \ell \in \mathbb{Z}} \, c_{-1} (n,\ell) \, q^n\, y^{\ell}, \;
\label{psimo}
\ee
is obtained using \eqref{etphexp} and the expansion $ 1/\eta^{24} (\sigma) = \sum_{j \geq -1} d(j) q^j$. For $\ell > 0$,
the coefficients $c_{-1} (n,\ell)$ are inferred from the expression
\bea
\left( \sum_{j \geq -1} d(j) q^j \right) \left( \sum_{k\geq 1}(2k+\ell)
  q^{k^2+\ell k}  \right) =  \sum_{\substack{n\geq 0, k\geq1,\\ n-k^2-k\ell \geq -1 }} (2k+\ell)\, d(n-k^2-k\ell) \, q^n  \;,
\eea
which results in 
\be
c_{-1} (n,\ell)= \sum_{\substack{ k\geq1,\\ n-k^2-k\ell \geq -1 }} 
 (2k+\ell)\, d(n-k^2-k\ell)  \;\;\;,\;\;\; n > 0 \;.
\ee
This has a wall crossing interpretation, as follows. Expressing $c_{-1} (n,\ell)$ in terms of the transformed
charge bilinears $m_{\gamma} = -1, \, n_{\gamma} = n - k^2 -k\ell \geq -1 , \,  \ell_\gamma =  2k+\ell > 0 $ as
\be
c_{-1} (n,\ell) = 
\sum_{\gamma\in \{ T^k,  \;  k \in \mathbb{N} \} } \ell_\gamma\, d(m_\gamma)d(n_\gamma) \;,
\label{cmogam}
\ee
the coefficients $c_{-1} (n,\ell) $ can be interpreted as arising due to the crossing of T-walls that are described by $\mathrm{PSL}(2,\mathbb{Z})$ matrices $\gamma$
of the form 
\begin{equation} 
	T^k = \begin{pmatrix}
		1 & k\\
		0 & 1
	\end{pmatrix} \;\;\;,\;\;\; k \geq 1 \;.
	\label{Twallk}
\end{equation}
Alternatively one could view \eqref{cmogam} as a wall-crossing derivation of $c_{-1}(n,\,\ell)$. One sees that \eqref{cmogam} is precisely
$d_*$ computed in \eqref{dstexp}.

Let us now return to the microstate degeneracies of single centre $\frac14$ BPS black holes, which are given in terms of the Fourier coefficients
$c_m^F (n,\ell)$, where we may restrict to $0 \leq \ell \leq m$, as discussed above, and where $n > 0$ to ensure that $\Delta = 4 mn - \ell^2 >0$.
The Fourier coefficients $c_m^F (n,\ell)$ are calculated in the $\mathcal{R}$-chamber,
by integrating $\psi_m^F$ along a path that satisfies
 ${\rm Im } v/{\rm Im} \sigma = - \ell/2m$. On the other hand, they 
are also encoded in the polar coefficients of $\psi_m^F$ \cite{Ferrari:2017msn,Chowdhury:2019mnb}. 
Thus, computing the latter in the $\mathcal{R}$-chamber
leads to a complete determination of the microstate degeneracies $d(m, n, \ell)$ of single centre $\frac14$ BPS black holes. 
The polar coefficients of $\psi_m^F$ can be expressed as \cite{Sen_2011,Chowdhury:2019mnb}
\be
c_m^F (n, \ell) = \sum_{\gamma \in W(m,n,\ell)} (-1)^{\ell_{\gamma} + 1} \, |\ell_{\gamma}| \, d(m_{\gamma}) \, d(n_{\gamma}) \;\;\;,\;\;\; 
\Delta = 4 mn - \ell^2 < 0\;,
\label{decay2c}
\ee
where each of the summands represents the contribution of a wall-crossing jump to the index $d(m,n,\ell)$.

As we showed in this paper, the set of walls $W(m,n,\ell)$
that are crossed can be taken to consist of those that are crossed when following the path associated with the continued fraction decomposition of $\ell/2m$.

The same reasoning applies to the contribution of wall-crossing jumps to $c_m^F (n, \ell)$ in the ${R}$-strip for $\Delta = 0$ decadent states.

Equation \eqref{decay2c} can be readily generalized to the study of 
negative discriminant states 
in CHL models. These models are obtained
by taking a $\mathbb{Z}_N$ orbifold of heterotic string theory on $T^6$ with $N=2, 3, 5, 7$. 
In a CHL model, the  microstate degeneracies of $\frac14$ BPS dyons in the twisted sector
are encoded in a Siegel modular form $\Phi_k$
(with the weight $k$ given by $k = 24/(N+1) - 2$) that is invariant
under the congruent subgroup $\Gamma_0(N) \subset \mathrm{PSL}(2, \mathbb{Z})$ \cite{Jatkar:2005bh}.  $\Phi_k$ admits the Fourier expansion
\be
\frac{1}{\Phi_k (\rho, \sigma, v)} = \sum_{m=-1}^{\infty} \psi_{-k,m} (\sigma, v) \, e^{2 \pi i m \rho} \;.
\label{phikpsi}
\ee
The  Fourier-Jacobi coefficient $\psi_{-k,-1}$ is given by \cite{Jatkar:2005bh}
\be
\psi_{-k,-1} (\sigma, v) =  f_2 (\sigma)  \, \frac{ \eta^6(\sigma)}{ \vartheta_1^2 (\sigma, v)} \;,
\ee
where 
\be
f_1(\sigma) =\left[  \eta (\sigma)^{k+2} \eta (N \sigma)^{k+2} \right]^{-1}= 
\sum_{j=-1}^{\infty} d_1(j) q^j \;\;\;,\;\;\; f_2(\sigma) = \sum_{r= -1/N }^{\infty} d_2(r) q^r = f_1(\sigma/N) \;.
\ee
These expansions reflect the fact that the charge bilinears $m$ and $n$ in these models satisfy the bounds $m \geq -1, \, n \geq -1/N$. 
For $\ell > 0$, the Fourier expansion of $\psi_{-k,-1}$  in the $\mathcal{R}$-chamber,
\be
\psi_{-k,-1} (\sigma,v) =  \sum_{n, \ell \in \mathbb{Z}} \, c_{-k,-1} (n,\ell) \, q^n\, y^{\ell} \;,
\label{psik1}
\ee
results in 
\be
c_{-k,-1} (n,\ell)= \sum_{\substack{ k\geq1,\\ n-k^2-k \ell \geq -1/N }} 
 (2k+\ell)\,  d_2(n-k^2-k\ell )  \;.
 \ee
Setting $m_{\gamma} = -1, \, n_{\gamma} = n - k^2 -k\ell \geq  -1/N, \,  \ell_\gamma =  2k+\ell > 0 $ and using $d_1(-1)=1$, this can be 
written as 
\be
c_{-k,-1} (n,\ell) = 
\sum_{\gamma\in \{ T^k,  \;  k \in \mathbb{N} \} } \ell_\gamma\, d_1(m_\gamma) \, d_2(n_\gamma) \;,
\label{cmogamchl}
\ee
which has a T-wall crossing interpretation as in \eqref{Twallk}.

The Fourier-Jacobi coefficients $\psi_{-k,m}$ in \eqref{phikpsi} with $m\geq 0$ can again be decomposed into a mock Jacobi
form $\psi_{-k,m}^F$ and a polar part $\psi_{-k,m}^P$ \cite{Dabholkar:2012nd, Bossard:2018rlt, Chattopadhyaya:2018xvg}. Taking the range\footnote{One can restrict the range of 
$\ell$ to be 
$0 \leq \ell \leq m$ for even weight Jacobi forms $\psi^F_{-k,m}$, i.e. for $N=2, 3, 4, 5$,
see \cite{Dabholkar:2012nd}.}
of $\ell$ to be $0 \leq \ell <  2m$, one again concludes
that the Fourier coefficients of $\psi_{-k,m}$ (with $m>0$) equal the Fourier coefficients of $\psi_{-k,m}^F$ in the $\mathcal{R}$-chamber.
 Then, assuming that there is a generalized Rademacher expansion of the mock Jacobi form 
 $\psi_{-k,m}^F$ for congruent subgroups $\Gamma_0(N)$, 
 the Fourier coefficients
 $c_{-k,m}^F (n,\ell)$ of $\psi_{-k,m}^F$ with $\Delta <0$  will determine
 the Fourier coefficients  $c_{-k,m}^F (n,\ell)$ with $\Delta > 0$.
 The analogue of \eqref{decay2c} for these CHL models 
reads \cite{Sen_2011}
\be
c_{-k,m}^F (n, \ell) = \sum_{\gamma \in W_N(m,n,\ell)} (-1)^{\ell_{\gamma} + 1} \, |\ell_{\gamma}| \, d_1(m_{\gamma}) \, d_2(n_{\gamma}) \;\;\;,\;\;\; \Delta = 4 mn - \ell^2 < 0\;.
\label{decay2cCHL}
\ee
Each summand represents again the contribution of a bound state that disappears when crossing 
a wall of marginal stability associated with the continued fraction decomposition of $\ell/2m$,
keeping however only those walls whose associated matrix $\gamma$ lies in the subset \eqref{setWN}.
As discussed above,
if the last wall that is crossed in this manner results in 
$m_{\gamma} = -1$, the operation of 
additional $T$ matrices needs to be taken into account. These are associated with the 
decomposition \eqref{cmogamchl} of the Fourier coefficients of $\psi_{-k,-1}$, by retaining
the contributions in \eqref{cmogamchl} with $n_{\gamma} \geq -1/N$. These contributions
are then to be added. A similar argument holds in the case of $n_{\gamma} = -1/N$, except that here we take into
account $U$ matrices.

{\it Thus, summarising, 
assuming that there is a generalized Rademacher expansion for congruent subgroups $\Gamma_0(N)$, 
the microstate degeneracies of a single centre $\frac14$ BPS black hole in (the twisted sector of) CHL models with $N=1, 2, 3, 5, 7$
are encoded in the continued fraction
decomposition of $\ell/2m \in \mathbb{Q}$ and in walls 
 corresponding to additional $T$ and $U$ matrices (c.f. \eqref{seq}) associated with the Fourier coefficients 
$c_{-k,-1} (n,\ell)$ of $\psi_{-k,-1}$.}

A similar reasoning applies to $\Delta = 0$ decadent states, as in the case of heterotic string theory on $T^6$.

We close with two comments in the case of heterotic string theory on $T^6$.
As shown in \cite{Sen_2011}, all the S-walls in the strip $s = -1$ can be mapped to the wall $v_2=0$ in $(\rho_2, \sigma_2, v_2)$-space,
and hence can be mapped to the $s=0$ T-wall of the 
polar part, 
\bea
 \sum_{m=-1}^{\infty} \psi_m^P (\sigma,v)  \, e^{2 \pi i m \rho}  &=& \frac{1}{ \eta^{24} (\sigma)} 
\sum_{m=-1}^{\infty} d(m) \,  {\cal A}_{2,m} (\sigma,v) \, e^{2 \pi i m \rho} \nonumber\\
&=& \frac{1}{ \eta^{24} (\sigma)} \sum_{s \in \mathbb{Z}} 
\sum_{m=-1}^{\infty} d(m) \,  
\, e^{2 \pi i m \rho} \, \frac{ q^{m s^2 + s} y^{2 m s +1}}{(1-q^s y)^2} \;.
\label{polargen}
\eea
Likewise, the S-walls in other strips will be mapped to corresponding T-walls of the 
polar part. Thus, the 
wall-crossing formula for S-walls is encoded in \eqref{polargen}.

The continued fraction of $\ell/2m \in \mathbb{Q}$ can be  represented in terms of a Stern-Brocot tree and can be viewed as
an `inverse discrete attractor flow' 
\cite{Cheng:2008fc}.
\\

\subsection*{Acknowledgements}
We would like to thank Sergey Alexandrov and Abhiram Kidambi 
for useful discussions. 
This work was partially
supported by FCT/Portugal through UID/MAT/04459/2019, 
UIDB/04459/2020
and through the LisMath PhD fellowship
PD/BD/135527/2018 (M. Rossell\'o). 

\appendix

\section{Integral binary quadratic forms}

By definition, an integral binary quadratic form is a homogenous quadratic polynomial
$f \in \mathbb{Z} [x,y]$ \cite{halter},
\be
f(x,y) = a x^2 + b   xy + c y^2 \;.
\label{fxy}
\ee
The associated discriminant $\Delta$ is
\bea
\Delta = 4 a c - b^2 \;.
\eea
Binary quadratic forms are divided into four types according to $\Delta$, as follows 
\cite{topology-numbers}:

\begin{enumerate}
\item $\Delta > 0$: $f$ is an elliptic quadratic form; 

\item $\Delta = 0$: $f$ is a parabolic quadratic form; 

\item $- \Delta > 0$, but not a square:  $f$ is a hyperbolic quadratic form; 

\item $- \Delta > 0$, and a square:  $f$ is $0$-hyperbolic.

\end{enumerate}

When $\Delta >0$, the binary quadratic form is called definite, whereas when  $\Delta <0$, 
the binary quadratic form is called indefinite.

Any integral binary quadratic form $f$ has a symmetric matrix $A$ associated to it,
\bea
A = 
\begin{pmatrix}
2a & b\\
b & 2c 
\end{pmatrix} \;\;\;,\;\;\; \det A = \Delta \;.
\eea
Two integral binary quadratic forms $f$ and $g$ are called equivalent \cite{halter}, $f \sim g$, if $g = \gamma \circ f $ for some matrix 
$\gamma \in \mathrm{SL}(2, \mathbb{Z})$, where $\gamma \circ f$ is defined by
\bea
\gamma \circ f &=& f(p x + q y, r x + s y) = \tfrac12  (x,y) \,  \gamma^T A \, \gamma \, 
\begin{pmatrix}
x \\
y
\end{pmatrix} \nonumber\\
&=& a_{\gamma} \, x^2 + b_{\gamma} \,  xy + c_{\gamma} \, y^2  \;\;\;,\;\;\; \gamma = 
\begin{pmatrix}
p & q \\
r & s
\end{pmatrix}
\in \mathrm{SL}(2, \mathbb{Z}) \;.
\label{gamcircf}
\eea
Here,
\bea
a_{\gamma} &=&  p^2 a + p r b + r^2 c = f(p, r) \;,
\nonumber\\
b_{\gamma} &=& 2 p q a + 2 s r c + (p s + q r ) b \;,
\nonumber\\
c_{\gamma} &=& q^2 a + q s b + s^2 c = f(q, s) \;.
\eea
The equivalence relation $f \sim g$ defines equivalence classes of  integral binary quadratic forms.
When a form $f$ is equivalent to itself under a transformation $\gamma$, i.e.  $f = \gamma \circ f $,
one speaks of self-equivalence of a form. Non-trivial  self-equivalences of a form are called automorphs
of the form \cite{buell1989binary}.

Binary quadratic forms have been shown to play a role in the study of BPS black holes \cite{Moore:1998zu,Moore:1998pn,Benjamin:2018mdo,Gunaydin:2019xxl,Banerjee:2020lxj,Borsten:2020nqi}. 
In this paper, we focused on the cases $\Delta < 0$ 
and $\Delta = 0$.
The associated integral binary quadratic forms are thus either hyperbolic/$0$-hyperbolic or parabolic.
In the case of parabolic forms, any parabolic form can brought to the form $a x^2$. Thus,
for each non-zero integer $a$ there is just one equivalence class of parabolic forms, with
 $a x^2$ being a representative in this equivalence class. In the case of hyperbolic forms,
 it can be shown that there are only a finite number of equivalence classes \cite{topology-numbers}.
 Properties of binary quadratic forms can be studied by their topographs \cite{conway,topology-numbers}.

In the following, let us consider indefinite integral binary quadratic forms,
and let us specify these for the various types of decadent dyons discussed in this paper.
First,  consider decadent dyons $(m, n, \ell)$ in $\mathbb{Z}_N$ CHL orbifold models with torsion $I=1$.
To these we associate 
the following integral binary quadratic form,
\be
f(x,y) = N m x^2 - N \ell xy + N n y^2 \;,
\label{fxyN}
\ee  
where we recall that for CHL models with  $N$ prime (with values $N=1, 2, 3, 5, 7$),
the charge bilinears are quantized according to 
\be
m=  \tfrac12 Q_m^2 \in \mathbb{Z} \;\;\;,\;\;\;  n=  \tfrac12 Q_e^2 \in \frac{1}{N} \mathbb{Z}  \;\;\;,\;\;\; \ell = Q_e \cdot Q_m \in \mathbb{Z}
\;.
\label{mnlN}
\ee
Subjecting $f$ to the $\mathrm{SL}(2, \mathbb{Z})$-transformation \eqref{gamcircf}, we obtain 
\bea
N m_{\gamma}  &=& N m p^2 - N \ell p r + N n r^2  \;, \nonumber\\
N n_{\gamma} &=& N m q^2 - N \ell q s   + N n s^2  \;, \nonumber\\
N \ell_{\gamma} &=& - 2 N m pq - 2 N n  r s  + N \ell (p s  + q r)  \;.
\eea
Requiring 
$m_{\gamma} , N n_{\gamma},  \ell_{\gamma} \in \mathbb{Z} $
restricts $\gamma$ to lie in the congruence subgroup $\Gamma_0(N) \subset \mathrm{PSL}(2, \mathbb{Z})$.

Next, we consider decadent dyons $(m, n, \ell)$ in the $N=1$ CHL model with arbitrary torsion $I$. 
This requires the charge bilinears to be quantized according to 
\bea
m \in  \mathbb{Z}  \;\;\;,\;\;\; n \in t^2 \mathbb{Z } \;\;\;,\;\;\;
\ell \in t  \mathbb{Z} \;,
\eea
with $t \vert I$.
To these dyons we associate 
the following integral binary quadratic form,
\be
f(x,y) = t^2 m x^2 - t \ell xy + n y^2 \;.
\label{fxyk}
\ee
Under a transformation $\gamma \in  \mathrm{SL}(2, \mathbb{Z})$-transformation, $(m,n,\ell)$ transform as
\bea
m_{\gamma}  &=& m p^2 -  \ell p r + n r^2 \;, \nonumber\\
n_{\gamma}  &=& m q^2 -  \ell q s   +  n s^2 \;, \nonumber\\
\ell_{\gamma} &=& - 2 m pq - 2 n   r s  + \ell (p s  + q r) \;.
\eea
Demanding  $m_{\gamma} \in  \mathbb{Z}, \; n_{\gamma} \in t^2 \mathbb{Z}, \; \ell_{\gamma} \in t  \mathbb{Z} $
implies
that $q$ is a multiple of $t$, and hence  $\gamma$  lies in the 
subgroup $\Gamma^0(t) \subset \mathrm{PSL}(2, \mathbb{Z})$.

We now 
consider the zeroes of $f(z,1)$ in \eqref{fxyk}, where 
$z = x/y$ (with $y \neq 0$), 
\bea
z_{\pm} = \frac{\ell}{2t m} \pm \frac{\sqrt{\ell^2 - 4 mn}}{2 t m} \;.
\label{zt}
\eea
In the range $z_- < z < z_+$, $f(z,1)$ attains its minimum at $z_0 = \tfrac12 (z_+ + z_- ) = \ell/2tm$.
Now observe that $t^2 m_{\gamma} = f(p, r) = r^2 f(p/r, 1) $ (here we assume $r \neq 0$). 
For $p/r$ in the range $z_- < p/r < z_+$ we have $m_{\gamma} < 0$.
Consider choosing a path in the $R$-strip that starts in the $\mathcal{R}$-chamber and moves down to negative values of $m_{\gamma}$,
ending at  $t^2 m_{\gamma} = f(p, r) < 0$.
Any rational value $z= p/r$ in the range $z_- < z < z_+$ will satisfies this. 
A natural choice is $z=z_0$, which is the value where $f$
attains its minimum.

The theory of indefinite binary quadratic forms naturally incorporates the use of continued
fractions via the link between Gauss' reduction algorithm and the continued fraction of $z_{\pm}$ \cite{halter}.
We now consider the case $t=1$. 
The continued fraction of $z_+$ and $z_-$ can yield charge bilinears of the form $(m_{\gamma}, n_{\gamma} = -1, \ell_{\gamma})$ and 
$(m_{\gamma} = -1, n_{\gamma}, \ell_{\gamma})$, respectively. The contributions of these states need to be identified under the electric/magnetic BSM phenomenon \cite{Sen_2011,Chowdhury:2012jq,Chowdhury:2019mnb}  in order to avoid overcounting.
The theory of indefinite binary quadratic forms also explains the appearance of the Brahmagupta-Pell equation $x^2  - | \Delta | \, y^2  = 4$
in the context
of the dyonic BSM phenomenon \cite{Chowdhury:2019mnb}, where all contributions of the form $(m_{\gamma} = -1, n_{\gamma} = -1, \ell_{\gamma})$
must be identified. This can be viewed as identifying automorphs of the indefinite binary quadratic form \eqref{fxyk} with $t=1$ and 
with $(m,n,\ell) = (-1, -1, \ell)$.
Automorphs of an indefinite binary quadratic form 
with non-square discriminant $-\Delta$
are in one-to-one correspondence with the solutions to the above Brahmagupta-Pell equation (see, for instance, Theorem 3.9 in \cite{buell1989binary}).

As we showed in this paper, the choice $z=z_0$ is 
a universal choice that works for all $m> 0$ and $\Delta \leq 0$, i.e. its continued fraction yields a set of decay walls that circumvent the phenomenon of BSM.


\providecommand{\href}[2]{#2}\begingroup\raggedright\endgroup

\end{document}